\documentclass[11pt,fleqn]{article}
\usepackage[utf8]{inputenc}
\usepackage{natbib}
\usepackage{tikz}
\usetikzlibrary{arrows}
\usepackage{xcolor} 
\usepackage{booktabs}
\usepackage{multirow}
\usepackage{hyperref}

\usepackage{amsmath}
\usepackage{graphicx}
\usepackage[a4paper, top=2.5cm, bottom=2.5cm, left=2.5cm, right=2.5cm]{geometry}

\usepackage{bbold}
\usepackage{todonotes}
\usepackage{amsthm}
\theoremstyle{definition}
\usetikzlibrary{tikzmark}
\usepackage{bm}
\usepackage{setspace}
\newcommand{\bA}{\bm A}

\newcommand{\bu}{\bm u}

\newcommand{\bw}{\bm w}

\newcommand{\bz}{\bm z}

\newcommand{\bdelta}{\bm \delta}

\newcommand{\boeta}{\bm \eta}

\usepackage{dsfont}
\usepackage{smartdiagram}
\usepackage{comment}
\usepackage{color}
\usepackage{url}
\usepackage{float}
\usepackage{xcolor}
\usepackage[ruled,vlined]{algorithm2e}
\usepackage[normalem]{ulem}
\usetikzlibrary{shapes,decorations,arrows,calc,arrows.meta,fit,positioning}
\tikzset{
    -Latex,auto,node distance =1 cm and 1 cm,semithick,
    state/.style ={ellipse, draw, minimum width = 0.7 cm},
    point/.style = {circle, draw, inner sep=0.04cm,fill,node contents={}},
    bidirected/.style={Latex-Latex,dashed},
    el/.style = {inner sep=2pt, align=left, sloped}
}

\title{
\LARGE{
High-Dimensional Granger Causality for Climatic Attribution}}
\author{Friedrich, M.$^{\mathsection}$, Margaritella, L.$^{\dagger}$\thanks{Corresponding author, email: \href{mailto:luca.margaritella@nek.lu.se}{luca.margaritella@nek.lu.se}. The second and third author thank NWO for financial support under grant number 452-17-010. Previous versions of this paper have been presented at the vEGU21 and (EC)$^2$ 2021 conferences and the Maastricht Workshop on Environmental Time Series Analysis. We gratefully acknowledge the comments by participants at these conferences. The datasets employed (.rds), as well as the R scripts to reproduce the analyses in this paper can be freely downloaded from the dedicated GitHub repository at \url{https://github.com/Marga8/Climate_Granger_Causality}.}, Smeekes, S.$^{\ddagger}$\\ \small{$^{\mathsection}$Vrije Universiteit Amsterdam, $^{\dagger}$Lund University, $^{\ddagger}$Maastricht University }}
\date{}
\begin{document}
\bibliographystyle{apalike2}
\maketitle
\begin{abstract}
In this paper we test for Granger causality in high-dimensional vector autoregressive models (VARs)
to disentangle and interpret the complex causal chains linking radiative forcings and global temperatures.~By allowing for high dimensionality in the model, we can enrich the information set with relevant natural and anthropogenic forcing variables to obtain reliable causal relations.~This provides a step forward from existing climatology literature, which has mostly treated these variables in isolation in small models. Additionally, our framework allows to disregard the order of integration of the variables by directly estimating the VAR in levels, thus avoiding accumulating biases coming from unit-root and cointegration tests. This is of particular appeal for climate time series which are well known to contain stochastic trends and long memory. We are thus able to establish causal networks linking radiative forcings to global temperatures and to connect radiative forcings among themselves, thereby allowing for tracing the path of dynamic causal effects through the system.
\end{abstract}

\maketitle

%
\doublespacing
\section{Introduction}\label{sec_intro}
While being originally introduced to investigate causal relationships in economics, the concept of Granger causality, dating back to \citet{granger1969investigating}, has developed into an established tool in studies of the climate system. 
Over the past 20 years, tests for Granger causality have gained increasing recognition in the climate literature as an important tool, proving to be more appropriate than lagged linear regression to infer causality \citep{mcgraw2018memory}. As a result, they have been widely applied across various fields of climate science.
For example, \citet{mosedale2006granger} study the relationship between sea surface temperatures and the North Atlantic Oscillation. \citet{jiang2015observational} test for Granger causal effects of vegetation change on climate in China, while \citet{kong2018vegetation} investigate the effects of climate factors on vegetation. More examples include \citet{bach2019local} with an application of Granger causality on sea surface temperatures and atmospheric variables, \citet{silva2021detecting} investigating global precipitation responses to El Ni\~{n}o surface temperature changes as well as \citet{chandio2022assessing} who study the effect of climate change on cereal production in Bangladesh. 

A considerable strand of the literature has focused specifically on the effects of human influences on the climate system, or ``climate change attribution'', with papers by \citet{kaufmann1997evidence}, \citet{Triacca2001granger}, \citet{triacca2005granger}, \citet{attanasio2011detecting}, \citet{attanasio2012contribution}, \citet{pasini2012evidence}, \citet{Triacca2013anthropogenic} and \citet{stern2014anthropogenic}, among the main references. These studies perform a Granger causal investigation between temperature and various radiative forcings which include natural as well as anthropogenic forcings. Notably, most of these previous studies apply tests for Granger causality in a framework with no more than two, sometimes three, variables at a time. Alternatively, as in \citet{bach2019local}, several atmospheric variables are combined by aggregating them into a broader measure, in order to use bivariate Granger causality tests. 

The necessity of limiting the number of variables in such analyses is statistical in nature, and arises because Granger causality relies on statistical tests to determine the significance of coefficients within a vector autoregressive (VAR) model. If the number of coefficients to estimate becomes too large, the data do not contain sufficient information to estimate
the parameters, and consequently standard statistical tools suffer from the ``curse of dimensionality'', resulting in estimators with high variance that overfit the data. However, by limiting the analysis to just a small set of variables, Granger causality might only reflect predictability, lacking robustness to spurious discoveries. As the climate is a complex system, it is impossible to condition the causal relation on all existing sources. Nonetheless, recent advancements in  high-dimensional statistics, such as those introduced in \citet{hecq2023granger}, at least make it as possible to perform the analysis in much more comprehensive and exhaustive networks. 

Our proposed approach employs active dimensionality reduction, thus enabling the estimation of VARs with many variabels, testing for Granger causality, and facilitating the interpretation of the outcomes. 
Another important feature in the proposed procedure is that no care is needed towards the time series properties of the variables considered. The data can therefore directly enter the model without pre-testing for unit roots and cointegration. This avoids the extra uncertainty that comes with pre-testing, and it is of great appeal for climate time series, which are well known to contain stochastic trends and long memory. 

Several alternative approaches to causality have recently been considered in climate research. These include structural causal models \citep{perezsuay2019causal}, conditional independence-based discovery algorithms \citep{runge2019inferring}, Bayesian VARs with impulse response analysis \citep{coulombe2021arctic} and multivariate Liang-Kleeman causality via the Kalman filter \citep{zhou2024estimating}. Each approach has its relative merits; we believe that our Granger causality  approach complements these existing ones.

Our dataset comprises various climate time series, coupled with GDP data, covering the time span from 1850 to 2018 with a yearly frequency. In particular, we obtain data on near surface temperature anomalies, various natural and anthropogenic forcings, as well as ocean heat content and the El Niño-Southern Oscillation index (ENSO). The inclusion of variables such as GDP, ocean heat content and ENSO seems especially new in the literature on climate change attribution.
We find Granger causal effects on temperature anomalies from several variables. Aggregate greenhouse gases (GHGs), ocean heat content, GDP and stratospheric aerosols cause temperature anomalies. The Granger causal relation between ENSO and temperature anomalies is bidirectional, and we also find that GDP has a Granger causal effect on ocean heat content. When we decompose GHGs into CO$_2$, CH$_4$ and N$_2$O, we discover a direct causal connection between CH$_4$ and temperature anomalies and several indirect paths from CO$_2$ and N$_2$O. The bidirectional causal link between ENSO and temperature is robust to this alternative model choice and likewise are our findings that ocean heat content and stratospheric aerosols cause temperature anomalies. We show how these findings are found to be robust to the choice of lag length and to the potential nonstationarity/cointegration of the series in our system.

The remainder of the paper is organized as follows. Section \ref{sec_methodology} discusses the methodology of Granger causality testing in high-dimensional VARs. Section \ref{sec_analysis} contains the main empirical analysis subdivided in two sub-sections: Section \ref{subsec_1} uses aggregated GHGs. Section \ref{subsec_3} disaggregates the GHGs into its three main gas components (CO$_2$, CH$_4$ and N$_2$O). Section \ref{sec_sensitivity} performs a sensitivity analysis on the lag-length specification. Finally, Section \ref{Sec_conclud_rmks} concludes.

\section{Granger Causality in Large Vector Autoregressions} \label{sec_methodology}

A variable $X$ is said to \emph{Granger cause} variable $Y$ if knowing the past of $X$ helps to improve the prediction of $Y$. Granger causality is a relative concept in the sense that the prediction including the past of $X$ needs to be measured against a ``benchmark'', that is, a prediction of $Y$ using a specific \emph{information set}. Let $\Omega_{t-1}$ denote this information set consisting of the history of selected variables up to time $t - 1$. Let $\mathcal{P}_t (Y|\Omega_{t-1})$ denote the best prediction of $Y$ at time $t$ given $\Omega_{t-1}$. Then we say that $X$ does not Granger cause $Y$ if $\mathcal{P}_t (Y|\Omega_{t-1}) = \mathcal{P}_t (Y|\Omega_{t-1}^{+X})$, where $\Omega_{t-1}^{+X}$ denotes the extended information set by including the past of $X$. In other words, $X$ does not Granger cause $Y$ relative to $\Omega_{t-1}$ if knowledge of the past of $X$ does not improve predictions given the knowledge of $\Omega_{t-1}$.

The composition of the information set plays a crucial role in the interpretation of the finding of Granger (non)causality. In the most basic version, which is often used in empirical studies, the information set consists only of the past of the variable $Y$. While finding Granger causality from $X$ to $Y$ still has a clear interpretation in terms of predictability, it is difficult to attach a causal meaning to such a result; for example, Granger causality will also be found if there is a third variable $W$ which causes both $X$ and $Y$. To control for its effect, we need to include it in the information set. Therefore, in order to give Granger causality a meaningful causal interpretation,\footnote{One caveat of Granger causality regardless of the size of the information set, is that it only considers dynamic relations where the effect does not materialize within the same time period but has a lagged influence. With Granger causality only remaining silent about contemporaneous relations, it can provide a partial description of causality at best. However common in many applications such as in climatology where often long run trends are of interest, the attribution of contemporaneous causality is of less importance; the key information is in the dynamics, which Granger causality captures.} one should make the information set as large as possible to include all possible confounders. Indeed, \citet{granger1969investigating} originally envisioned the information set to ``be all the information in the universe” (p. 428). This is in stark contrast to typical applications of Granger causality analysis which involve at most a handful of control variables \citep[see a.o.,][for examples in climate sciences]{Triacca2013anthropogenic,kaufmann2013stochastic}; the reason generally being the difficulty of estimating statistical models with many variables due to the ``curse of dimensionality''. While conditioning on all the information in the universe clearly remains an infeasible concept, advances in statistical learning and high-dimensional methods have made working with much larger information sets possible.

To move towards a practical implementation of a high dimensional Granger causality test, we first need to operationalize the abstract best predictor $\mathcal{P}_t (Y|\Omega_{t-1})$. We focus on prediction in mean and linear models. As we are mostly interested in trends in our application, the focus on the mean is natural. While the focus on linear models may be up for discussion, one should realize that we do not necessarily require the linear model to be the truth, but only require it to be a decent approximation in terms of the underlying Granger causal relations. In addition, assuming linearity greatly helps in dealing with the high-dimensionality, which would become considerably more complicated if more flexible nonlinear models are allowed for.

Let $\bz_t$ denote the $K$-dimensional system of time series under consideration. Following convention, we order the causing (`treatment') variable $x_t$ first, outcome variable $y_t$ second, and the $K - 2$ controls $\bw_t = (w_{1,t,} \ldots, w_{K-2,t})'$ third, such that we can write $\bz_t = (x_t, y_t, \bw_{t}')'$. We assume that the dynamic system $\bz_t$ can be represented as a vector autoregressive process of order $p$ (VAR($p$)): 
\begin{equation}\label{var}
\bz_{t}=\bA_{1} \bz_{t-1}+\cdots+\bA_{p} \bz_{t-p}+\bu_{t}, \quad t=p+1, \ldots, T
\end{equation}
Here $\bA_1,\ldots,\bA_p$ are the transition matrices that determine how the lags of $y_t$ affect its current values, and $u_t$ is the serially uncorrelated (but potentially cross-correlated) vector of error terms.

In this setup, Granger causality is concerned with the parameters in the transition matrices $\bA_1,\ldots,\bA_p$; it is easy to see that $x_t$ does not Granger cause $y_t$ if and only if the $(2,1)$-th element in $\bA_j$ -- corresponding to the impact of $x_{t-j}$ on $y_t$ -- is equal to zero for all $j=1,\ldots,p$. In order to test for Granger (non)causality, we can then consider the regression of $y_t$ on the lags of all variables:
\begin{equation}\label{eq:regr_GC}
y_t = \sum_{j=1}^p \beta_j x_{t-j} + \sum_{j=1}^p \delta_{y,j} y_{t-j} + \sum_{j=1}^p \bdelta_{w,j}' \bw_{t-j} + u_{y,t},
\end{equation}
and test whether $\beta_1 = \ldots = \beta_p = 0$.

Note that this regression quickly becomes high-dimensional. With just $K=6$ variables and $p=5$ lags, we would have to estimate 30 parameters. If one had annual data for, say, 50 years, estimation and inference would become very challenging due to the ``curse of dimensionality''. In order to be able to estimate such high-dimensional systems, we must impose some structure on the parameters. A common assumption is that of \textit{sparsity}, which states that many parameters in \eqref{eq:regr_GC} are (close to) zero, such that the regression can be well approximated with a simpler model, containing only the most important variables. Importantly, one does not need to assume knowledge on which variables are important; statistical learning techniques, such as the lasso \citep{tibshirani1996regression}, can determine the important variables in a purely data-driven way.

\cite{hecq2021granger} propose a test for Granger causality under the setting of sparsity, utilising the post-double-selection framework of \cite{belloni2014high}. In this framework, for the outcome variable $y_t$ and the treatment variables $x_{t-1}, \ldots, x_{t-p}$, high-dimensional regression models are estimated on all variables using the lasso ($p+1$ lasso regressions in total). All variables that are deemed to be important in at least one of the regressions, are then collected and added, alongside the treatment variables, as explanatory variables in a final regression estimated by ordinary least squares (OLS). By performing variable selection based on two criteria, we minimise the probability of missing a variable, which would cause omitted variable bias and therefore endogeneity. The two criteria are: relevance to explain the outcome variable, and correlation to the treatment variables.  \cite{hecq2021granger} show that this procedure leads to valid inference and asymptotically normal OLS estimators in the final regression, such that standard Wald or LM tests can be used.

However, \cite{hecq2021granger} work on the assumption that the vector autoregressive model is stationary; if the time series under consideration contain stochastic trends, such that they need to be differenced in order to become stationary, standard inference breaks down, even in low dimensions \citep{granger1974spurious}. One could conduct tests for unit roots and difference the series accordingly to achieve stationarity; this will, however, delete information about long-run co-movements. Given that this is arguably the most relevant information for climatic attribution, this strategy is not appropriate for our application. In addition, there is disagreement in the literature as to whether climate variables, such as temperature, are stationary around a (broken) linear trend or contain a stochastic trend. We refer the interested reader to  \citet{kaufmann2010stochastic}, \citet{estrada2010reply} and \citet{kaufmann2013stochastic} for an overview of this discussion. In our approach, we do not have to perform any pretesting to decide on the stationarity properties of the variables involved in our model. Therefore, this disagreement does not add any additional uncertainty in our analysis.

Instead, one can apply the lag-augmentation strategy proposed by \cite{toda1995statistical}, in which redundant lags are added to the model. This can account for the stochastic trends, by providing an automatic differencing mechanism. This strategy, applied in the climate context by a.o., \citet{triacca2005granger}, restores asymptotic normality of the estimators, regardless of the order of integration and the presence of stochastic trends. It is, however, not directly applicable in high dimensions. To illustrate why, consider the former example with $K=6$ variables and $p=5$ lags. If one has to add redundant lags such as $p=6$ or $p=7$, this clearly only makes the ``curse of dimensionality'' worse.

\cite{hecq2023granger} develop a new method that combines the high-dimensional Granger causality test with an appropriate lag-augmentation strategy, and show that this method retains asymptotic normality in high-dimensional sparse models even in the presence of stochastic trends.

The final procedure is straightforward to implement. First, consider the high-dimensional regressions
\begin{equation} \label{eq:stage1}
\begin{split}
y_t &= \sum_{j=1}^p \beta_j x_{t-j} + \sum_{j=1}^p \delta_{y,j} y_{t-j} + \sum_{j=1}^p \bdelta_{w,j}' \bw_{t-j} + u_{y,t},\\
x_{t-i} &= \sum_{j=1, j\neq i}^p \eta_{x,j} x_{t-j} + \sum_{j=1}^p \eta_{y,j} y_{t-j} + \sum_{j=1}^p \boeta_{w,j}' \bw_{t-j} + e_{j,t}, \qquad i = 1, \ldots, p.
\end{split}
\end{equation}
We estimate these with the lasso, leaving $x_{t-1}, \ldots, x_{t-p}$ and $y_{t-1}, \ldots, y_{t-p}$ unpenalised, such that these are always included to avoid spurious regression. We then collect all selected variables from $\bw_{t-1}, \ldots, \bw_{t-p}$ in each regression; let $\bw_{t-1}^*, \ldots, \bw_{t-p}^*$ denote the subset of variables that have been selected in at least one regression. 

We then estimate the final regression
\begin{equation}\label{eq_post_selection}
y_t = \sum_{j=1}^{p+d} \beta_j x_{t-j} + \sum_{j=1}^p \delta_{y,j} y_{t-j} + \sum_{j=1}^p \bdelta_{w,j}^{*\prime} \bw_{t-j}^* + u_{y,t}
\end{equation}
by OLS, and test whether $\beta_1 = \ldots = \beta_p = 0$.\footnote{Following \cite{hecq2023granger} we use the LM test, although the Wald test can equally well be used. Both have a standard chi-squared asymptotic distribution with $p$ degrees of freedom under the null hypothesis of no Granger causality.} Here $d$ represents the maximum order of integration that we believe the variables in the system to have; in other words, $d$ represents how often we need to difference the data to achieve stationarity and remove the trends. In most applications, including ours, $d=2$ suffices. Note that the actual order of integration may be lower, but not higher, in order for the test to remain valid. Note also how, in contrast to the original \citet{toda1995statistical} procedure, the lag-augmentation $d$ is only applied to one variable, i.e., the Granger causing $x_t$ and not on all the regressors. This circumvents the worsening of the ``curse of dimensionality'', allowing for high dimensional applications. 

In principle, any statistical learning method could be used to estimate the first stage regressions in \eqref{eq:stage1}; we follow \cite{hecq2023granger} in using the lasso with the tuning parameter selected by the Bayesian information criterion. The final choice that remains to be made is to select the original lag length $p$. A too low value of $p$ should be avoided since some of the $d$ additional lags of the lag augmentation would, in case of a too low $p$, be needed to model the dynamics, and the procedure will not work. However, selecting $p$ too large will not have a major detrimental effect on the test. With this in mind, \cite{hecq2023granger} provide a simple data-driven way to select $p$ as an `informed upper bound' on the true lag length, based on applying information criteria to individual autoregressions. We follow their approach, but also investigate the robustness of the results to different values of $p$.\footnote{For further discussion of the details of the implementation, as well as a theoretical and simulation analysis of the performance of the Granger causality test, we refer to \cite{hecq2023granger}.}

\section{Empirical Granger Causality Analysis}\label{sec_analysis}
\subsection{Aggregated Greenhouse Gases Analysis}\label{subsec_1}
\begin{figure}
\centering
\includegraphics[width=\textwidth]{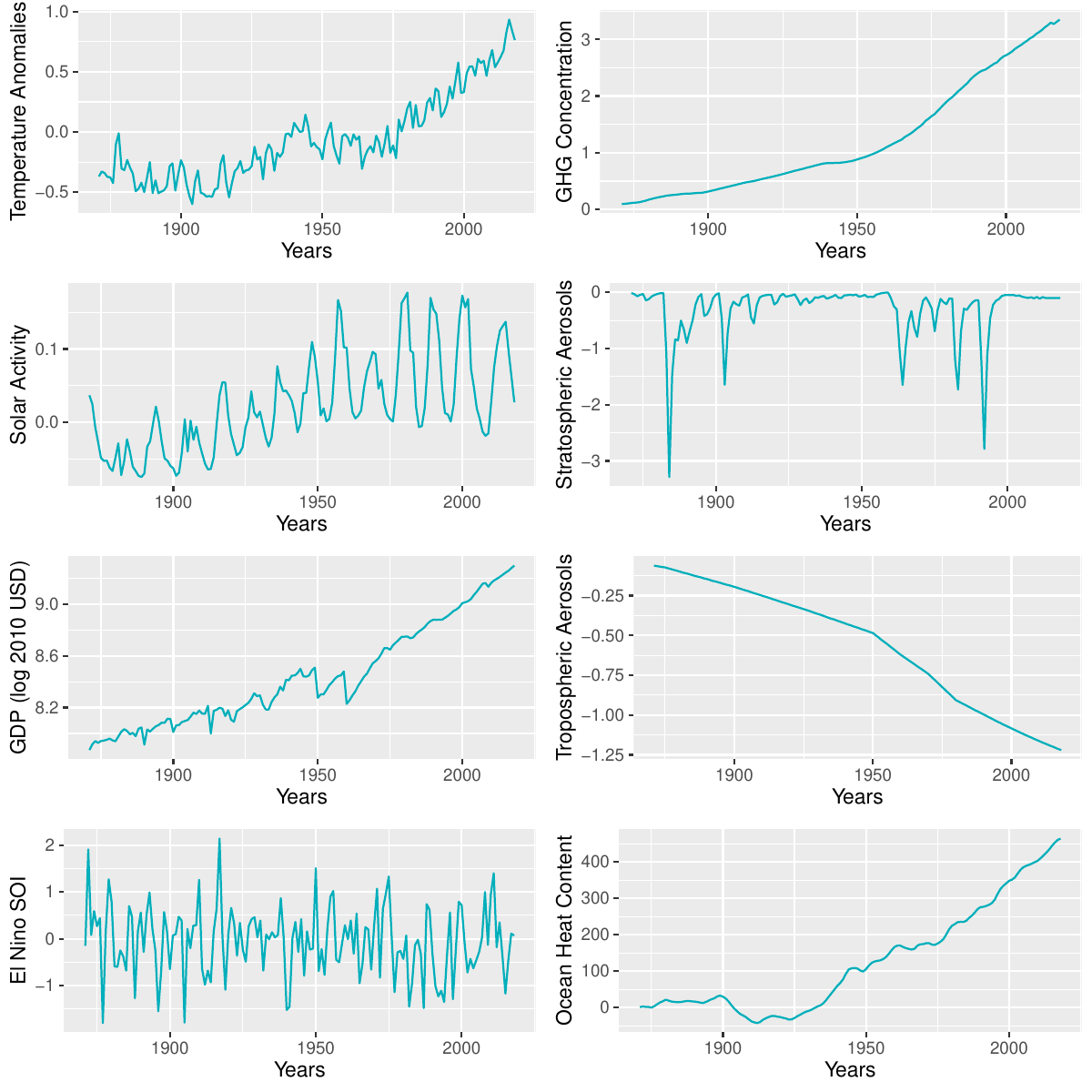}
\caption{Individual time series plots of our collected data set of annual observations from 1871 to 2018.}\label{fig_miller_series}
\end{figure}

We employ annual time series data ranging from $1871$ to $2018$. Below, we provide details on the considered variables, including their abbreviations (in bold), measurement units (in square brackets)\footnote{The abbreviation ``Fe'' stands for ``\emph{effective forcings}''.}, and sources. These series are plotted in Figure \ref{fig_miller_series}.
\begin{itemize}
    \item \textbf{T}: Temperature Anomaly [°C]. Source: \citet{morice2020updated}.
    \item \textbf{G}: Greenhouse gases [Fe($W/m^2$)]. Source: \citet{hansen2017young}.
    \item \textbf{S}: Solar Activity [Fe($W/m^2$)]. Source: \citet{hansen2017young}.
    \item \textbf{V}: Stratospheric Aerosols from Volcanic Activity [Fe($W/m^2$)]. Source: \citet{hansen2017young}.
    \item \textbf{A}: Tropospheric Aerosols and Surface Albedo [Fe($W/m^2$)]. Source: \citet{hansen2017young}.
    \item \textbf{Y}: GDP [log 2010 US\$]. Source: Maddison Project Database 2020 (\citealp{bolt2013maddison}) for $1850-1959$, World Bank data for $1960-2018$.
     \item \textbf{N}: El Niño–Southern Oscillation index (ENSO) [°C]. Source: Climate Research Unit (CRU) at the University of East Anglia\footnote{\url{https://crudata.uea.ac.uk/cru/data/soi/} SOI calculations are based on the method given by \citet{ropelewski1987extension}.} 
    \item \textbf{O}: Ocean Heat Content [$10^{21}$J, full depth]. Source: \citet{zanna2019global}\footnote{\url{https://laurezanna.github.io/post/ohc_pnas_dataset/}}.
\end{itemize}

This set of variables considers key factors of the climate evolution. \emph{Effective forcings} of $ G, S, V, A$ are considered according to \citet{hansen2017young}, thus removing the effect of rapid adjustments occurring in the atmosphere which do not relate to longer term surface temperature responses. First, we consider temperature anomalies, which is the natural target to be explained in climate change attribution models. From its displayed time series in the top left panel of Figure \ref{fig_miller_series} a pronounced increase after 1900 is clearly visible. Second, we consider an aggregated measure of GHGs as in \citet{hansen2017young}. We consider individual, disaggregated time series in Part \ref{subsec_3} of this section. In the aggregated series displayed in the top right panel in Figure \ref{fig_miller_series}, we observe a steady upward trend which has been accelerated since the 1950s. Third, solar activity. This has an important, purely natural influence on the climate system, and in our data displays a seasonal and overall slightly upward trending pattern. Fourth, we consider aerosols, which can also be distinguished by being of natural and of anthropogenic source. The former come from e.g., volcanic eruptions, evaporation of seawater, hydrocarbon emissions from forested areas. The latter comes mostly from fuel combustion (diesel and biomass burning produce \emph{black aerosols} absorbing sun's energy) and burning of high-sulfur coal. In our data, stratospheric aerosols display several negative peaks but no overall trend while tropospheric aerosols follow a downward trend. As an addition to the literature, we also include global GDP which is steadily upward trending. Finally, we add two further climate variables: the El Niño–Southern Oscillation index (ENSO) and the Ocean heat content (OHC).\footnote{For the calculation of ENSO we refer to \url{https://www.ncdc.noaa.gov/teleconnections/enso/indicators/soi/}}   
Oscillations of annual CO$_2$ growth are correlated with global temperature and with the El Niño/La Niña cycle \citep[see][]{hansen2017young}.
\citet{thompson2008large} shows that El Niño is a natural source of variability responsible for the global warmth of $1939$–$1945$ and strong El Niño events have also occurred in $1997$–$1998$ and $2015$–$2016$ that might have boosted the temperature. Ocean heat content also needs to be accounted for in the analysis. In fact, the ocean has lots of thermal inertia and it might take up to centuries before the Earth surface temperature reaches most of its fast-feedback response to a change in climate forcing \citep[][]{hansen1985climate}. We can see in the bottom two panels of Figure \ref{fig_miller_series} that our ENSO data are fluctuating around a mean of zero while OHC shows a rapid upward trend since the 1920s.

\begin{figure}[ht]
    \centering
    \begin{minipage}{.5\textwidth}
        \centering
\includegraphics[width=0.95\textwidth, trim = {2.5cm 2.5cm 2cm 2cm},clip]{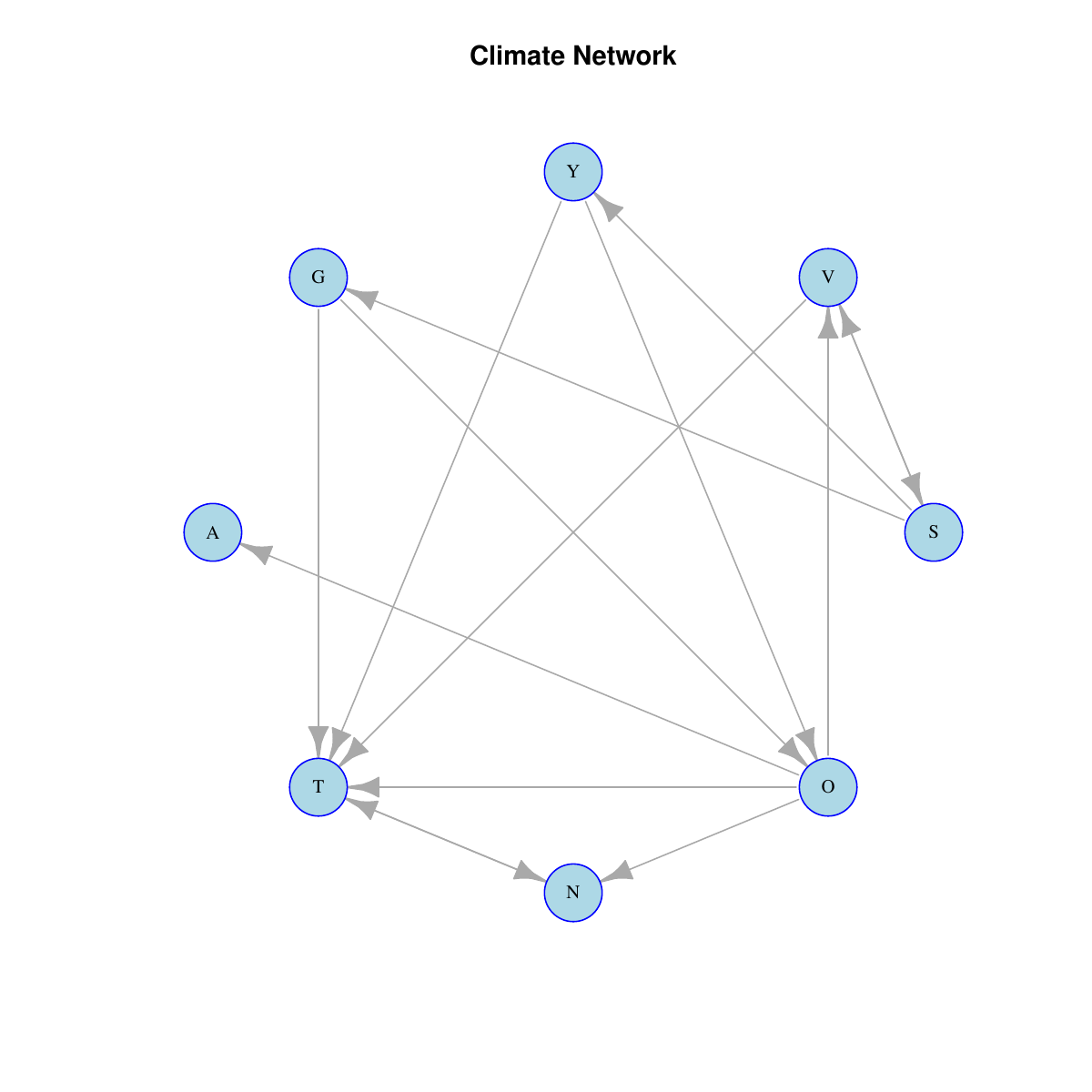}
\caption{Climate Network, $\alpha=0.1$, $p=3$}
\label{fig_miller_4}
\end{minipage}%
\begin{minipage}{0.5\textwidth}
\centering
\includegraphics[width=0.95\textwidth]{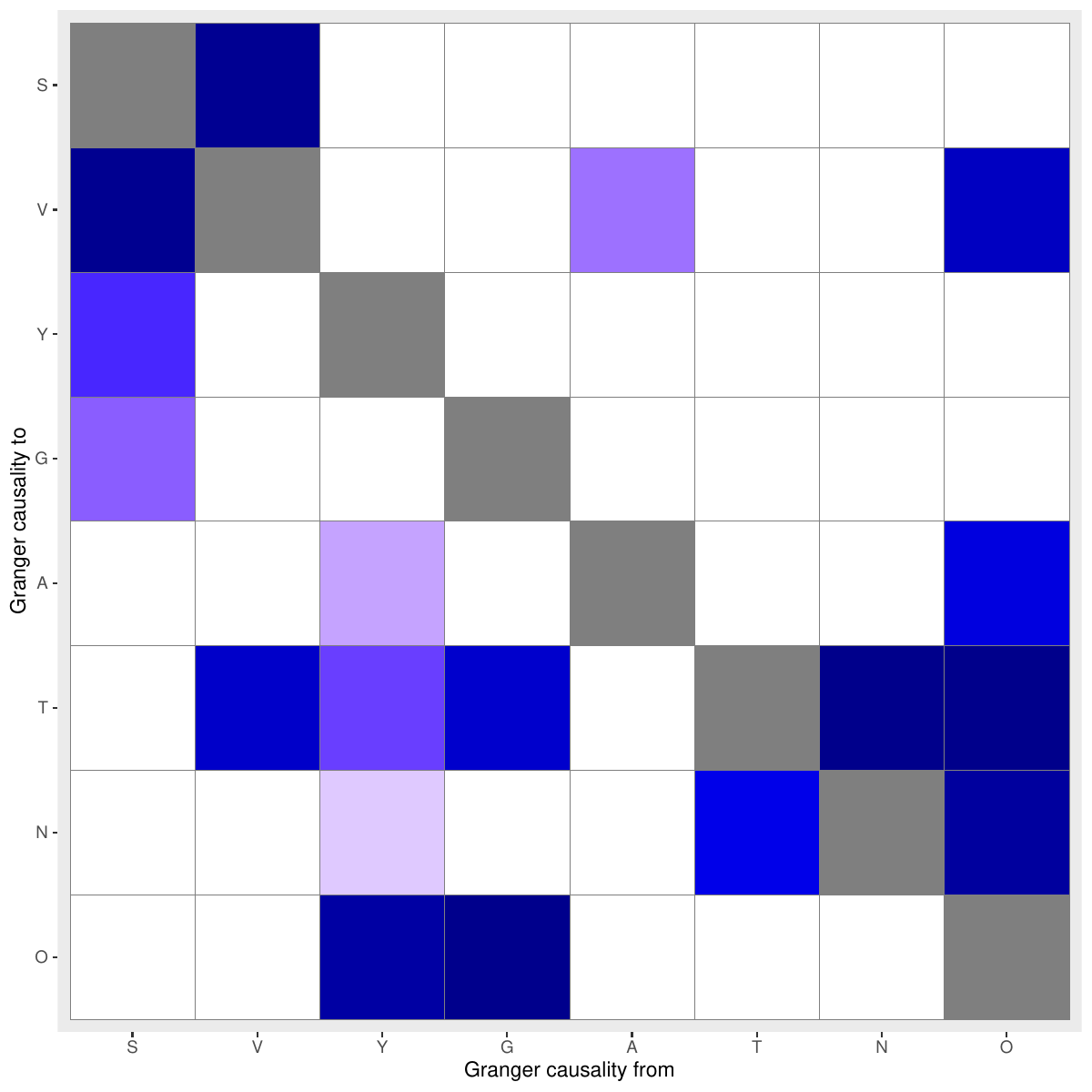}
\caption{ $p$-values heat-map}\label{fig_miller_5}
\end{minipage}
\end{figure}

Let us stack all the above series in a VAR model as in equation (\ref{var}). We use the methods described in Section \ref{sec_methodology} in order to obtain the causal network displayed in Figure \ref{fig_miller_4}. This graph plots all bivariate, conditional Granger causal relationships among the variables described above. The arrows represent a (directional) Granger causal link that was found at a significance level of 10\%.\footnote{All the analyses reported in the following sections have been carried out using \emph{R} \citep{R}. Scripts and data can be freely downloaded from the dedicated GitHub repository. See the ``Data Availability Statement".} 

\begin{figure}[ht]
    \centering
    \begin{minipage}{.5\textwidth}
        \centering
\includegraphics[width=0.9\textwidth, trim = {2.5cm 2.5cm 2cm 2cm},clip]{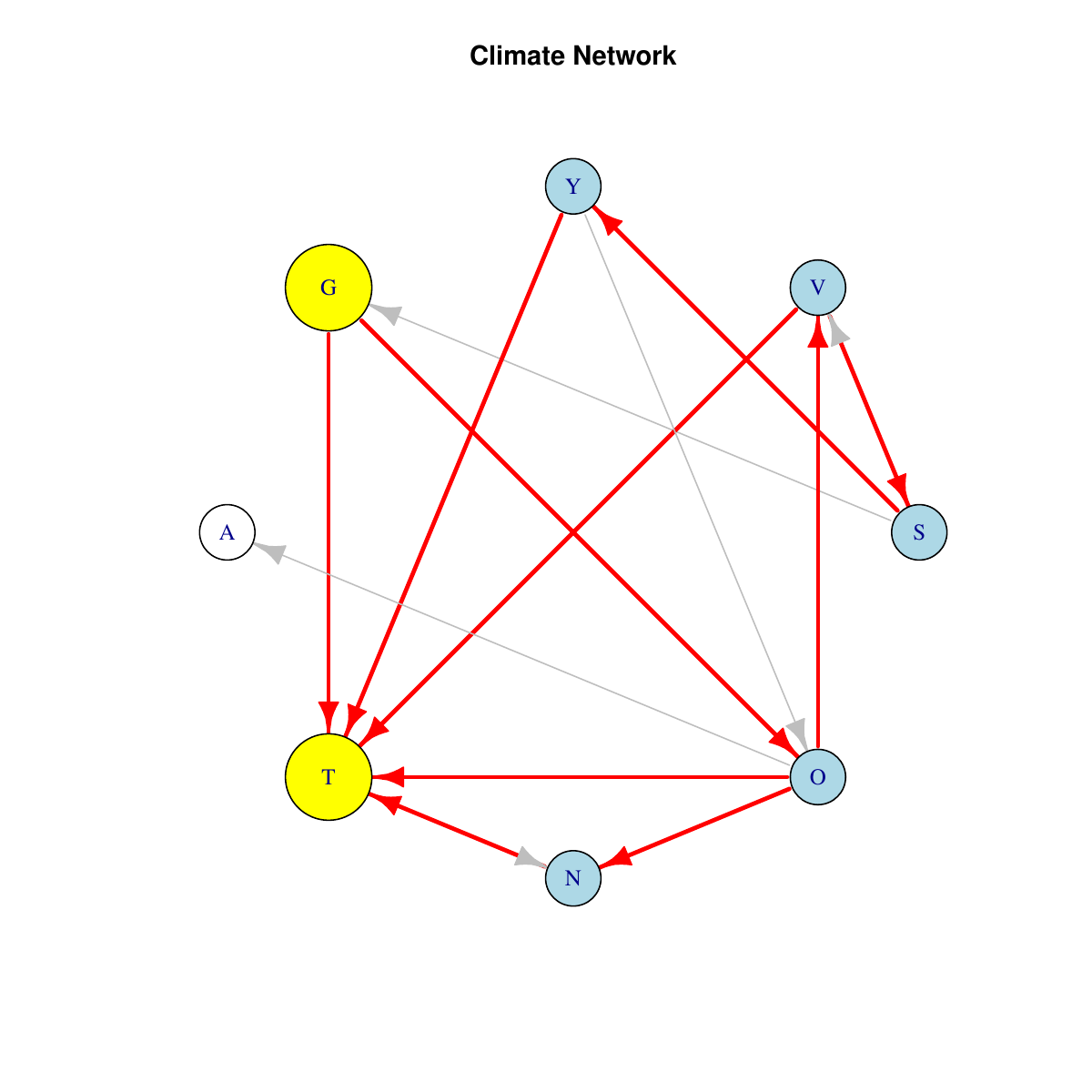}
\caption{$G$ to $T$ paths (5)}
\label{fig_GtoT}
\end{minipage}%
\begin{minipage}{0.5\textwidth}
\centering
\includegraphics[width=0.9\textwidth, trim = {2.5cm 2.5cm 2cm 2cm},clip]{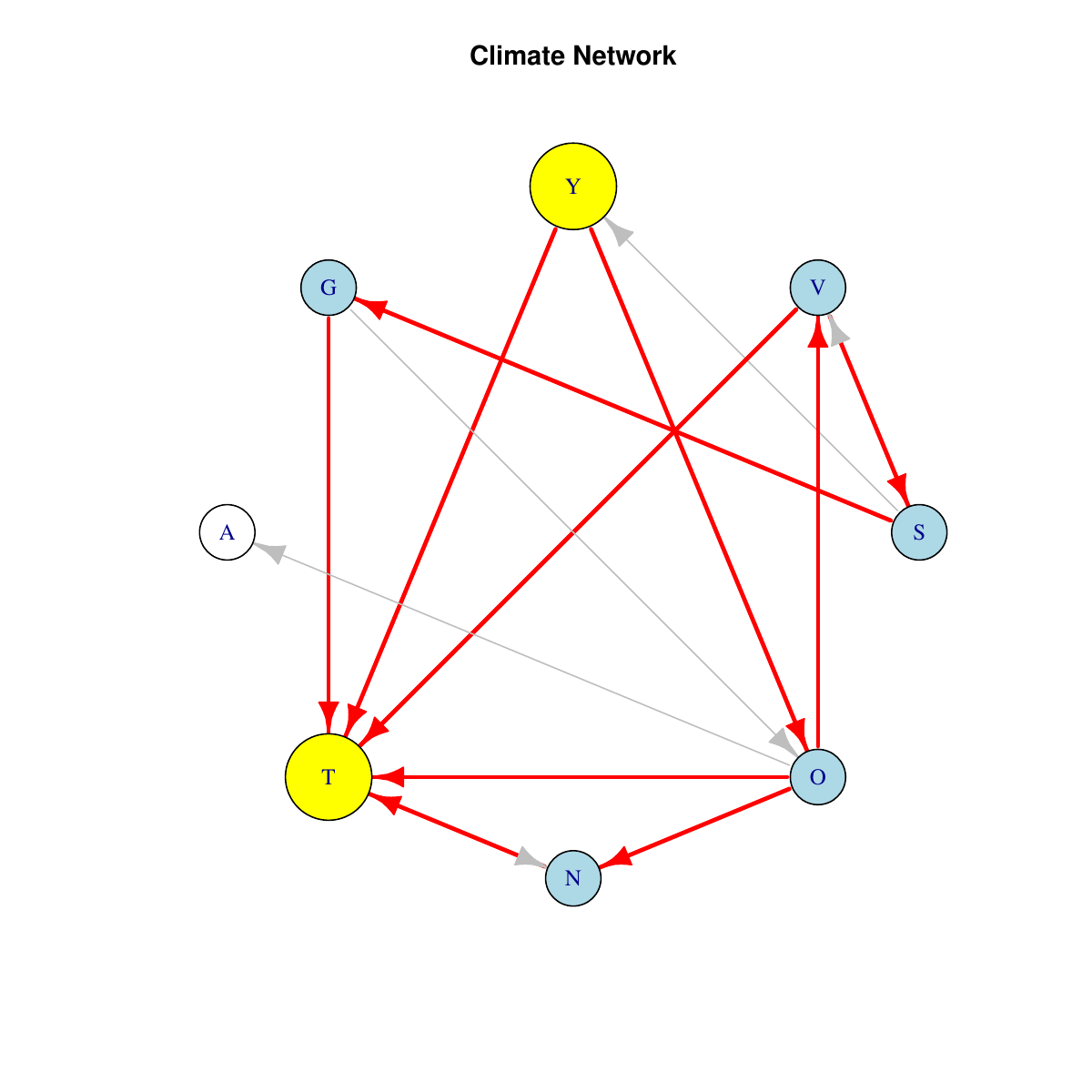}
\caption{ $Y$ to $T$ paths (5)}\label{fig_YtoT}
\end{minipage}
\end{figure}

Figure \ref{fig_miller_5} presents the corresponding heat map for the strength of the $p$-values of the Granger causality test for the different combinations among the considered series. The darker the color, the smaller the $p$-value. All white boxes are $p$-values larger than $15\%$.  
The results of this Granger causal network give us some key insights. Conditional on all other variables, we find evidence 
that greenhouse gas forcings ($G$) Granger causes temperature anomalies ($T$). This means that given our included natural forcing as well as production variables, anthropogenic forcings from GHGs hold predictive power with respect to temperature. However, temperature does not Granger cause anthropogenic forcings in the same system. This supports the findings of \citet{stern2014anthropogenic} who also find a unidirectional link from anthropogenic forcings to temperature in the presence of natural forcings. We additionally find a strong link 
to temperature from the following variables: ENSO ($N$), ocean heat content ($O$), stratospheric aerosols ($V$) and GDP ($Y$). The link between stratospheric aerosols and temperature has also been found in  \citet{stern2014anthropogenic}, while ocean heat content is taken as a purely exogenous variable in their paper. Our results indicate that part of the effect of GHGs on temperature might run through the ocean which takes up part of the increase in heat. A causal link between ENSO and temperature has not been tested before in the literature. However, \citet{stern2014anthropogenic} discuss the option of removing the effect of ENSO on the temperature and modelling the adjusted series. The authors, however, argue that these oscillations are an endogenous part of the climate system and therefore should not be removed. We therefore explicitly consider ENSO in our model and we find a causal link to temperature which runs in both directions. In addition, we find that GDP has a strong effect on Ocean Heat Content and that Solar activity causes Stratospheric aerosols as well as GHGs. 
\par From Figure \ref{fig_miller_4} we observe two \emph{feedback} relations: stratospheric aerosols with solar activity, and temperature with ENSO. The latter is of particular interest as observed in \citet{houghton2001climate}. 
An El Niño event is characterized by positive temperature anomalies in the eastern equatorial Pacific.~This reduces the sea surface temperature difference across the tropical Pacific.~As a consequence, the trade winds from east to west near the equator are weakened and the Southern Oscillation Index becomes anomalously negative, letting sea level to fall in the west and rise in the east by almost $25$ cm. At the same time, these weakened winds reduce the rise of cold water in the eastern equatorial Pacific, thereby strengthening the initial positive temperature anomaly. Thus, ENSO influences tropical climate but also possess a global influence: during and following El Niño, the global mean surface temperature increases as the ocean transfers heat to the atmosphere \citep[see][]{sun1998coordinated}.

From Figure \ref{fig_miller_4} we can also identify \emph{cycles}. GHGs ($G$) are found to lead ocean heat content ($O$), which in turn leads stratospheric aerosols ($V$) which is itself found to feedback to GHGs passing first through solar activity ($S$). The connection between $G$ and $O$ can be understood as follows: as GHGs act as a blocking layer trapping more energy from the sun, the oceans are absorbing more heat as a consequence, and this results in an increase in sea surface temperatures and rise of sea level. The connection between $O$ and $V$ is also readily justified. Even though we do not find a direct feedback of $V$ on $O$, there is a cyclic relation among the two, outsourced by solar activity and GHGs.
\citet{church2005significant} observe how large volcanic eruptions, emitting aerosols in large quantities, result in rapid reductions in ocean heat content and global mean sea level. They bring the example of the eruption of Mount Pinatubo, a stratovolcano in the Philippines, estimating as a consequence of the eruption a reduction in ocean heat content of about $3^*10^{22}$J and a global sea-level fall of about $5$ mm. Over the three years (coinciding with our lag-length) following such an eruption, the estimated decrease in evaporation is of up to $0.1$ mm day$^{-1}$. 

\par Focusing on the two arguably most interesting connections -- GHGs to temperature and GDP to temperature -- we also plot all possible causal paths for these two relationships in Figures \ref{fig_GtoT} and \ref{fig_YtoT}. These show that there is not only a direct link but the effect can also be indirect by going through (multiple) other variables. 
For greenhouse gases we identify five \emph{simple}\footnote{In graph theory a path is simple if the vertices it visits are not visited more than once.} Granger causal paths: one direct and four indirect, running through all the other nodes except for tropospheric aereosols ($A$). The shortest path larger than two nodes is $G\to O\to T$.
Interestingly, we also observe a causal path from GHGs passing through GDP. This is the longest path from GHGs, passing through ocean heat content, which in turn affects stratospheric aereosols, solar activity, GDP and finally temperature anomalies ($G\to O\to V\to S\to Y\to T$).
For GDP to temperature anomalies, we also observe a total of five causal paths among which one direct and four indirect passing through the same nodes as for the GHGs. The shortest path is analogous to the shortest path from $G$. Namely, we find $Y\to O\to T$. Interestingly, the longest path from $Y$ to $T$ passes through nodes in the exact same order as $G$ does. Namely, we observe a path $G\to O\to V\to S\to Y\to T$ and one $Y\to O\to V\to S\to G\to T$. This is a sensible finding: global production inevitably emits GHGs and their effect on other climate variables is therefore consequential. What stands out is that GHGs have, although indirectly, an effect on GDP too, and the two indirectly feedback to each others.

\subsection{Disaggregated Greenhouse Gases Analysis}\label{subsec_3}
\begin{figure}
\centering
\includegraphics[width=0.55\textwidth, trim = {0cm 13.5cm 10cm 0cm},clip]{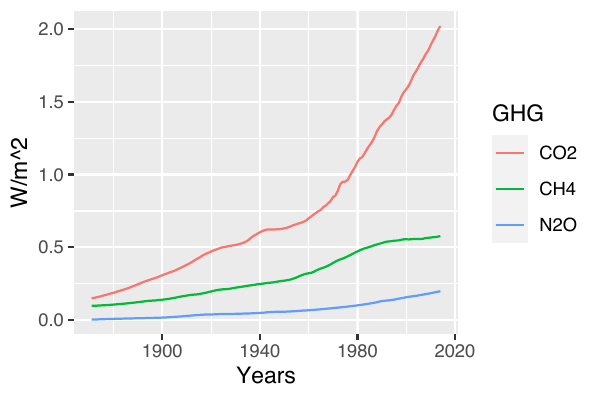}
\caption{Disaggregated GHGs series: the red line plots CO$_2$, the green line CH$_4$ and the blue line N$_2$O.}\label{ghg_series}
\end{figure}

In the previous setting we considered an aggregated measure of greenhouse gases ($G$) which is meant to represent the combined forcings of all the different anthropogenic emissions. Now instead, we consider the three main GHGs in terms of global warming potential: carbon dioxide ($CO_2$), methane ($CH_4$), nitrous oxide ($N_2O$). We use historical reconstructions as computed in \citet{meinshausen2017historical}\footnote{Data available at \url{https://www.climatecollege.unimelb.edu.au/cmip6}} to disentangle the single effects of the main GHGs on temperature. As the data are given in concentration as parts-per-billion (ppb), we transform them into radiative forcings by using the transformations from \citet{Hansen1998Climate} as shown in Table \ref{tab_forcings}.
\begin{table}
\caption{ Radiative Forcings Conversions Formulae}\label{tab_forcings}
\centering
\begin{tabular}{llc}
\midrule\midrule
 \textbf{Variable} & \textbf{Radiative forcing} & \textbf{Pre-ind. concentration}  \\\midrule
 \multirow{2}{*}{CO$_2$} & $F = f(c)-f(c_0)$ &  \multirow{2}{*}{$c_0\approx 280$ppm} \\
 & where $f(c) = 5.04\ln\left[c + 0.0005c^2\right]$ &   \\\midrule
 \multirow{2}{*}{CH$_4$} &  $F=0.04 \left(\sqrt{m}- \sqrt{m_0}\right)-\left[g(m, n_0)- g(m_0, n_0)\right]$ & \multirow{2}{*}{$m_0\approx 700$ppb}\\
 & where $g(m, n) = 0.5 \ln\left[1 + 0.00002(mn)^{0.75}\right]$ &\\\midrule
 N$_2$O & $F=0.14 (\sqrt{n}-\sqrt{n_0})-\left[g(m_0, n)-g(m_0, n_0)\right]$ & $n_0\approx 275$ppb \\
 \midrule\midrule
\end{tabular}
\end{table}
We integrate the three time series in place of G in the previous model set up, obtaining a total of $10$ series spanning the timeframe from $1871$ to $2014$. Figure \ref{ghg_series} displays the time series of the three main GHGs which are all upward trending, with $CO_2$ showing the most pronounced trend.

\begin{figure}
    \centering
    \begin{minipage}{.5\textwidth}
        \centering
\includegraphics[width=0.95\textwidth, trim = {2.5cm 2.5cm 2cm 2cm},clip]{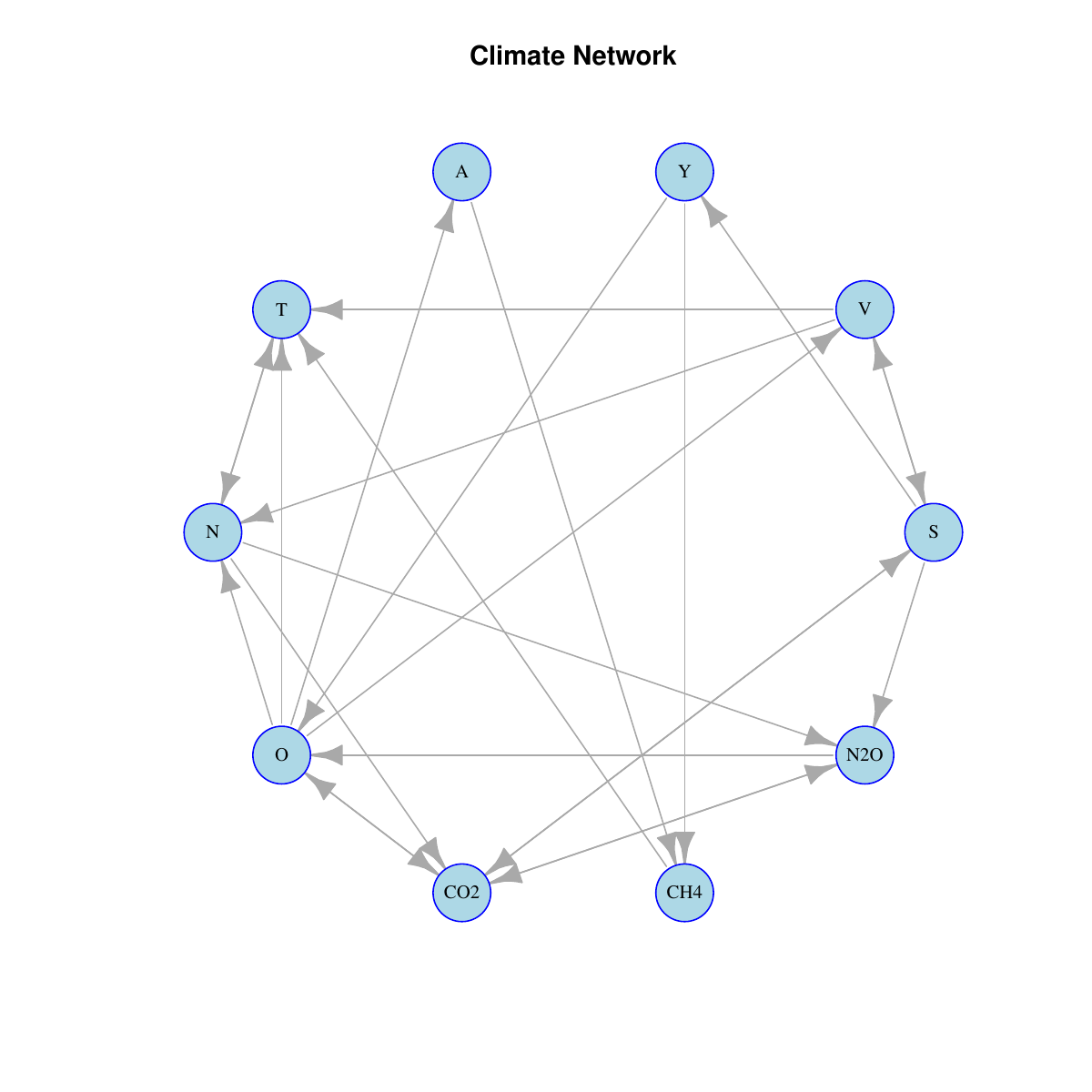}
\caption{Climate Network, $\alpha=0.1$, $p=3$}
\label{fig_disaggr1}
\end{minipage}%
\begin{minipage}{0.5\textwidth}
\centering
\includegraphics[width=0.95\textwidth]{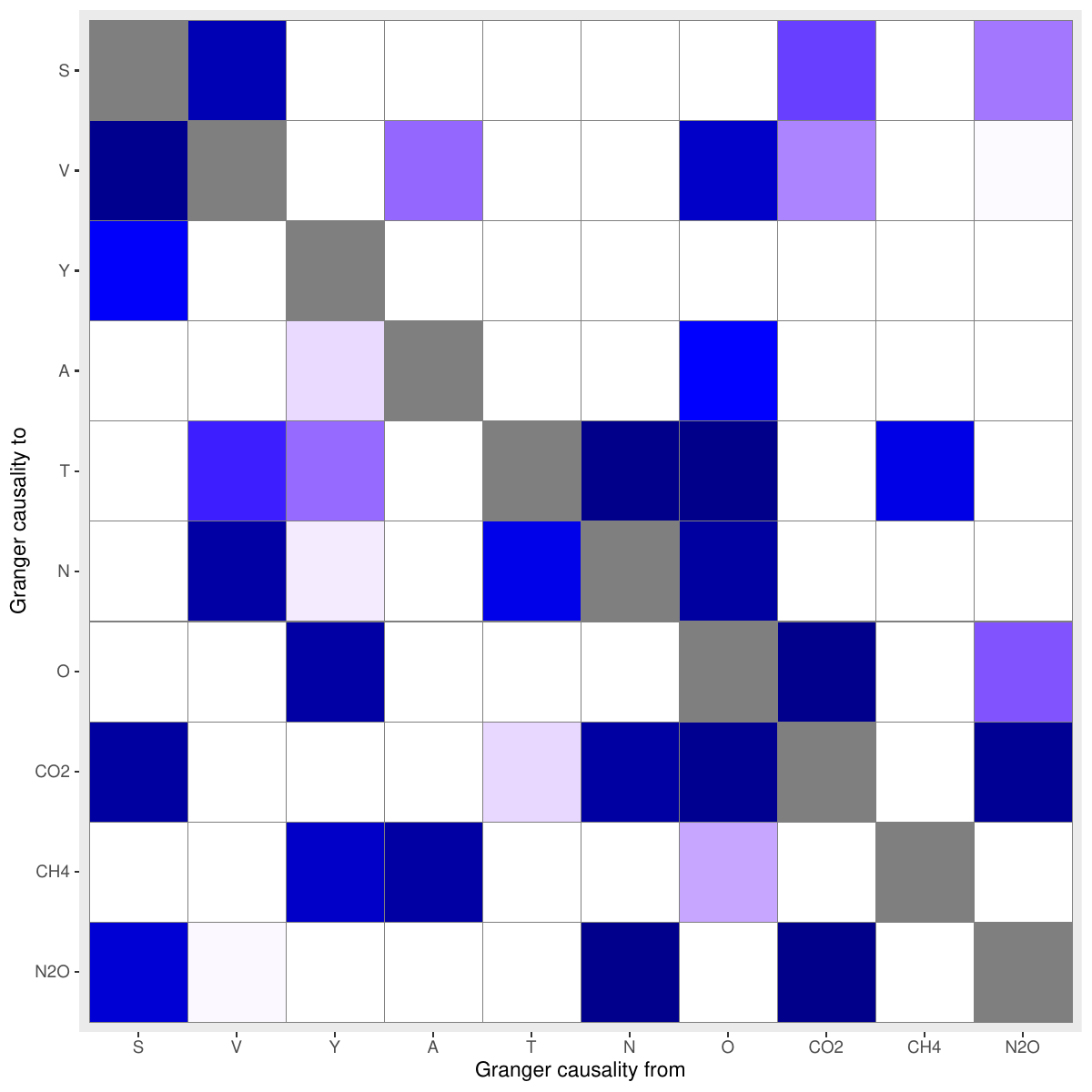}
\caption{$p$-values heat-map}\label{fig_disaggr2}
\end{minipage}
\end{figure}

\begin{figure}
    \centering
    \begin{minipage}{.325\textwidth}
        \centering
\includegraphics[width=0.95\textwidth, trim = {2.5cm 2.5cm 2cm 2cm},clip]{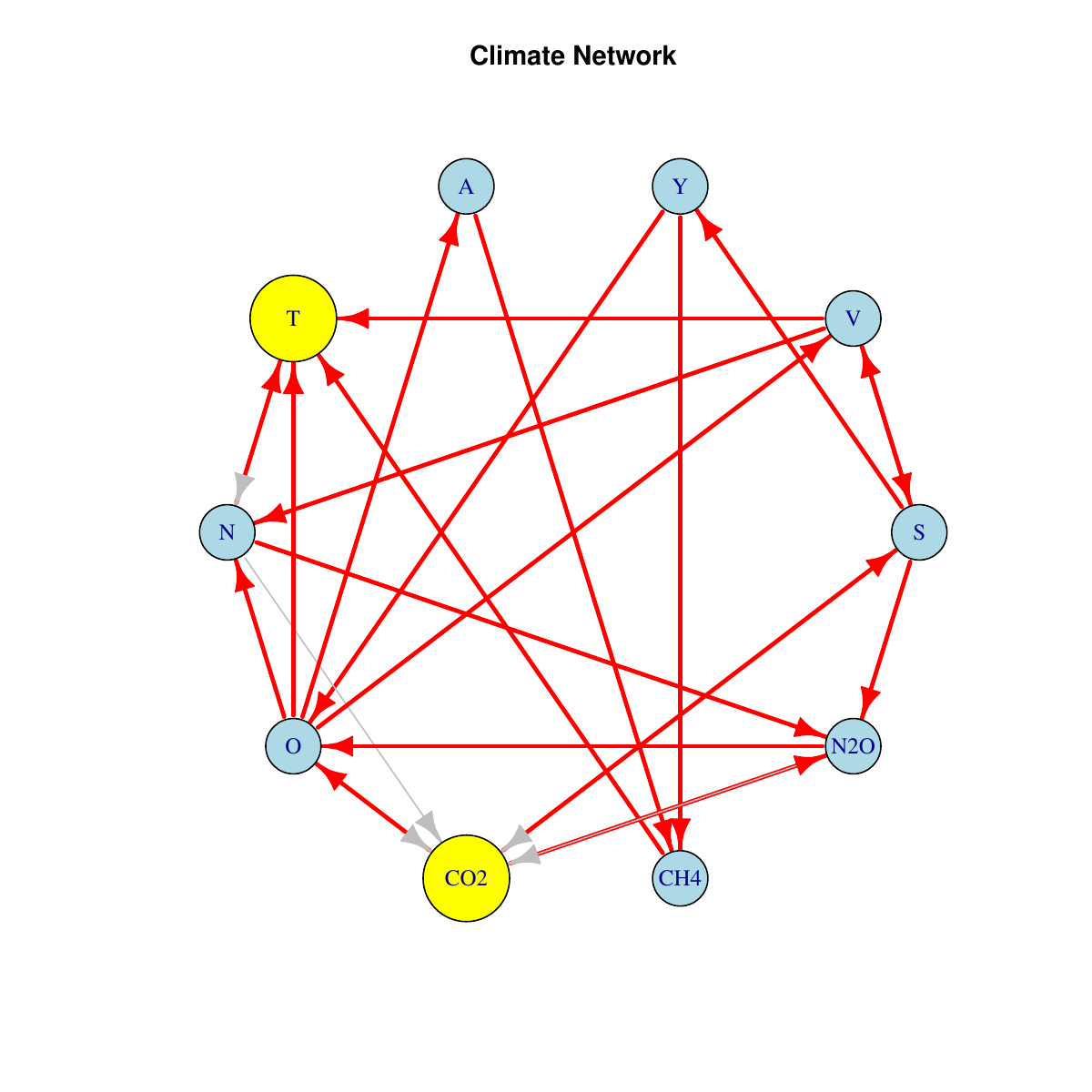}
\caption{CO$_2$ to T paths (27)}
\label{fig_disaggr3}
\end{minipage}%
\begin{minipage}{0.325\textwidth}
\centering
\includegraphics[width=0.95\textwidth, trim = {2.5cm 2.5cm 2cm 2cm},clip]{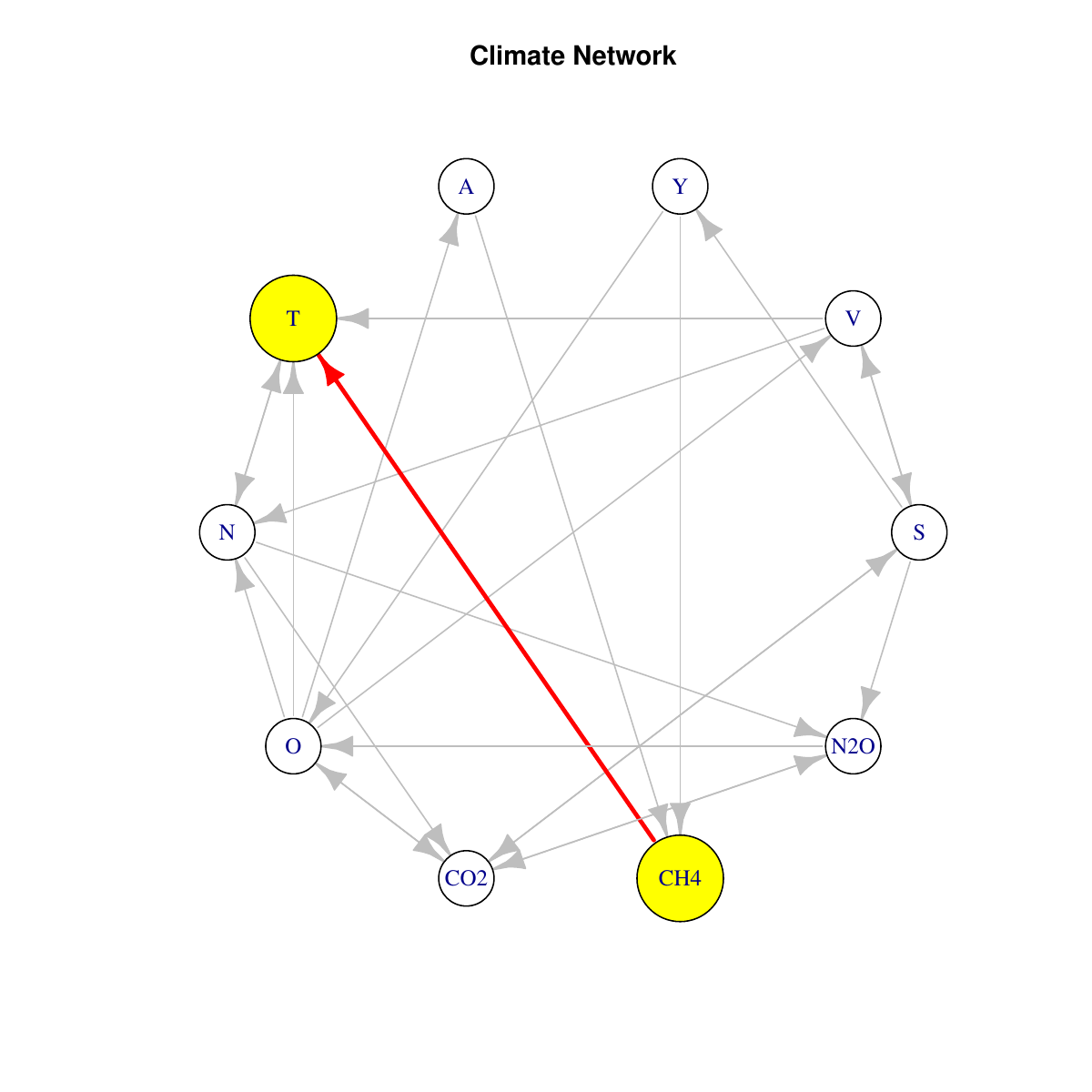}
\caption{CH$_4$ to T paths (1)}\label{fig_disaggr45}
\end{minipage}
\begin{minipage}{0.325\textwidth}
\centering
\includegraphics[width=0.95\textwidth, trim = {2.5cm 2.5cm 2cm 2cm},clip]{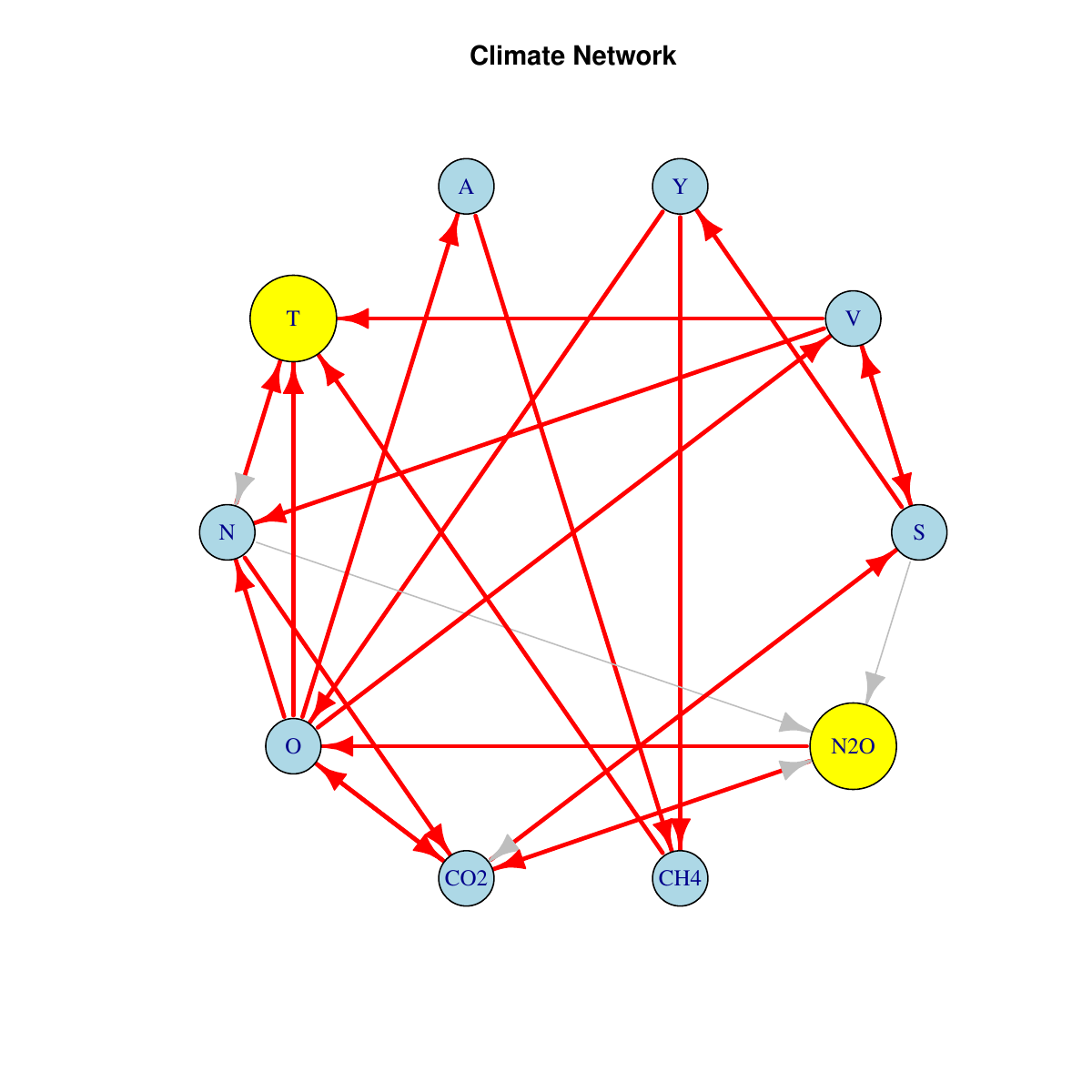}
\caption{ N$_2$O to T paths (26)}\label{fig_disaggr4}
\end{minipage}
\begin{minipage}{0.325\textwidth}
\centering
    \centering
    \includegraphics[width=0.95\textwidth, trim = {2.5cm 2.5cm 2cm 2cm},clip]{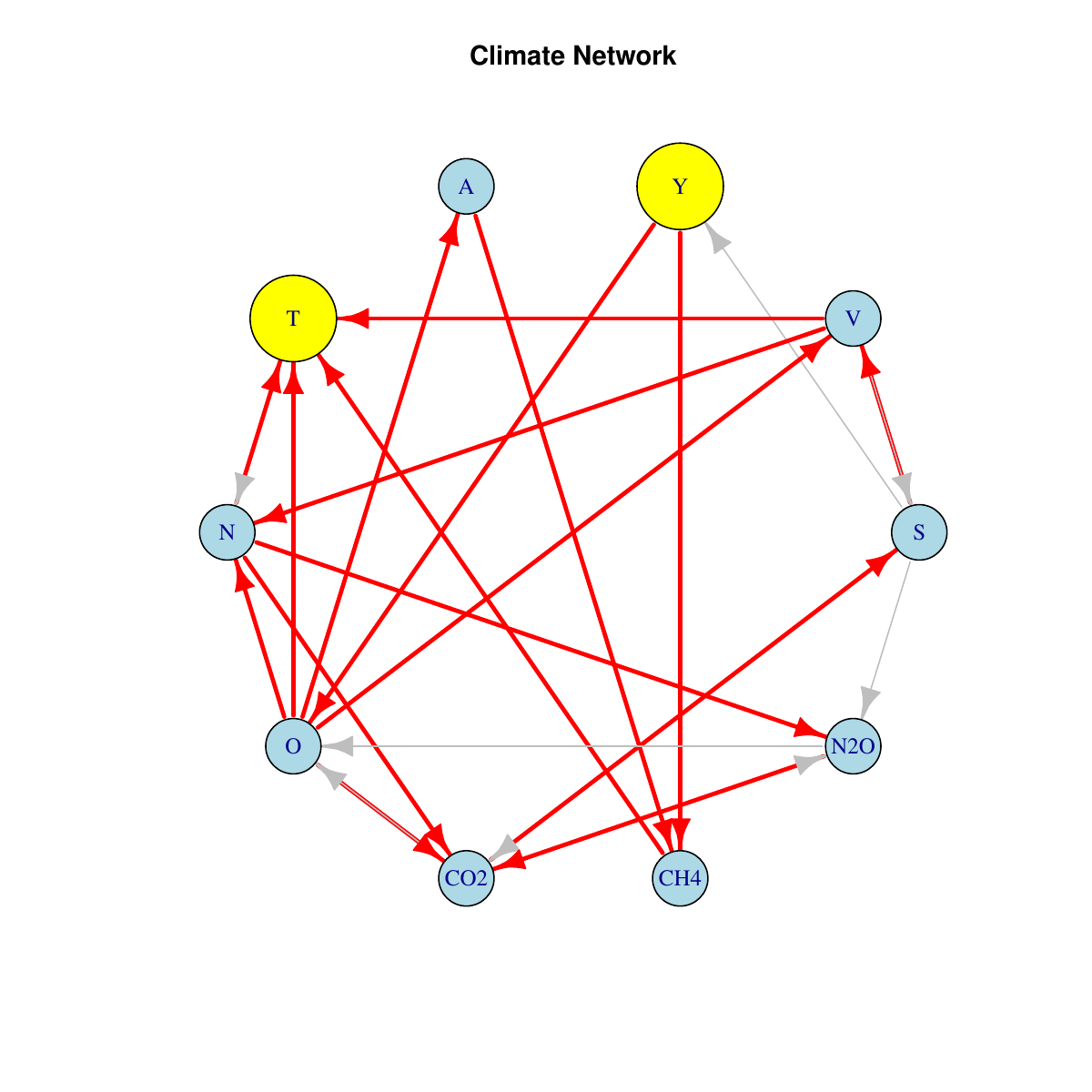}
    \caption{Y to T paths (10)}
    \label{Y to T disagr}
\end{minipage}

\end{figure}

Figures \ref{fig_disaggr1}-\ref{Y to T disagr} display the results. From Figure \ref{fig_disaggr1} we observe a total of 25 edges and that among the greenhouse gases considered, only $CH_4$ displays a direct connection to $T$.
 However, many indirect relations are observed from both $CO_2$ and $N_2O$ to $T$. In Figures \ref{fig_disaggr3}-\ref{fig_disaggr4} we highlight the amount of possible causal paths between, in turn, $CO_2$, $CH_4$, $N_2O$ and temperature anomalies. 
 We observe a total of $27$ paths from $CO_2$ to $T$ and $26$ from $N_2O$ to $T$. This shows how potentially intricate the effect of $CO_2$ and $N_2O$ on temperature is. The shortest path observed between $CO_2$ and $T$ is $CO_2\to O\to T$, while the longest touches on $7$ of the remaining vertices: $CO_2\to S\to V\to N\to N_2O\to O\to A\to CH_4\to T$. Likewise for $N_2O$, the shortest path observed is $N_2O\to O\to T$ and the longest touches again on $7$ of the remaining vertices as: $N_2O\to O\to V\to N\to CO_2\to S\to Y\to CH_4\to T$.  
 Interestingly, we observe a bidirectional, direct causal relation between $CO_2$ and $N_2O$, but not between $CO_2\leftrightarrow NH_4$ and $N_2O\leftrightarrow NH_4$.

 Finally, Figure \ref{Y to T disagr} displays the Granger causal paths from GDP to temperature. We observe a total of 10 paths passing through all the variables, including the greenhouse gases. This gives evidence of how GDP has a broad  interplay through time with several climate variables in the span of 3 years. Interestingly, there are two shortest paths from $Y$ to $T$, one passing through $CH_4$ and one, like in the case of $CO_2$ and $CH_4$, passing through ocean heat content. Instead, the longest path touches on $6$ of the remaining vertices, including two of the greenhouse gases: $Y\to O\to N\to N_2O\to CO_2\to S\to V\to T.$ In general, we observe temperature being caused directly from $N$, $O$, $CH_4$, $V$, where $N$, $O$ and $V$ are the same direct ones observed as in the analysis in Section \ref{subsec_1}, although here no direct effect of GDP is observed but instead methane is found to directly cause temperature anomalies.
 \par From Figure \ref{fig_disaggr1} we observe few \emph{feedback} relations too: as observed in Section \ref{subsec_1} a bidirectional relationship is established between ENSO and temperature. ENSO is the only variable for which temperature anomalies is found to be a leading indicator. We also observe a feedback between $CO_2$ and $N_2O$, one between $CO_2$ and ocean heat content, one between $CO_2$ and solar activity, one between $N_2O$ and solar activity and finally one between solar activity and stratospheric aerosols.
 
 Many more loops are now observed compared to Section \ref{subsec_1}, well demonstrating the complex, broad interplay among variables in climate systems. Even though no Eulerian cycle\footnote{In Graph Theory an Eulerian cycle is a paths that starts and ends at the same vertex, visiting all other vertices in its route.} is observed, almost all the variables in the system, except $A$, $T$, $CH_4$ have paths starting and ending in their own vertex. The variables with the most amount of observed cycles are solar activity ($S$) and stratospheric aerosols ($V$), respectively with $63$ and $60$ cycles. Vice-versa, the observed variable with the least amount of cycles is GDP, with only $8$.  
 For cycles starting respectively from $CO_2$ and $N_2O$ let us consider now respectively the shortest and longest cyclic (simple) paths. For $CO_2$, the shortest cyclic path runs through: $O,T,N,N_2O$. The longest instead are two, passing through: $S,Y,O,A,CH_4,T,N,N_2O$ or $O,V,S,Y,CH_4,T,N,N_2O$. Likewise, the min/max cyclical path observed from N$_2$O are respectively: $T,N,CO_2$ and $O,V,S,Y,CH_4,T,N,CO_2$.

What stands out from these cycles effects is that ENSO ($N$) is a crucial variable to account for in analyzing a climate system. In fact, ENSO appears in all the circular causal chains starting from greenhouse gases and passing through temperature as well as in all other $8$ unmentioned cycles passing by $T$. Ocean heat content ($O$) follows closely as it appears in $9/11$ of the cycles. Solar activity ($S$) appears in $8/11$, stratospheric aerosols ($V$) appears in $5/11$ while tropospheric aerosols ($A$) only appears in two cycles.

\par In the following Section \ref{sec_sensitivity} we are going to perform a sensitivity analysis on the lag-length $p$. One potential reason why we do not observe direct connections between greenhouse gases and temperature could be the slow response that temperature has to changing greenhouse gases. It might therefore be beneficial to manually enlarge the lag-length considered to verify these relations.

\section{Sensitivity Analysis: lag-length}\label{sec_sensitivity}

In the previous section, we employed a data-driven selection of the lag-length $p$. Its estimation is crucial in a VAR context and even more so in this lag-augmented version since the additional lag-augmentation $d$ depends on $p$ (see Section \ref{sec_methodology}). Here, we are interested in observing what happens to the aggregated greenhouse gas network and the disaggragated greenhouse gas network as in Figure \ref{fig_miller_4} and \ref{fig_disaggr1}, when the lag-length is manually augmented. This is interesting as, on the one hand, it shows how our modeling framework is able to handle larger dimensionalities. On the other hand, it is known that temperature exhibits a slow response to changes in many climate variables, including greenhouse gases. As such, considering larger lag-lengths can potentially uncover new connections. We previously estimated $p=3$, meaning the coefficients we reported in Section \ref{subsec_3} were expressing a compound effect of the relevant Granger causing variable to the Granger caused in the span of the past 3 years. We now consider the same analysis for a sequence of lag-lengths $p=(10, 30)$. This means that we have up to $302$ variables to estimate per equation for a given sample size of $144$ data points, thus, we need to exploit the high-dimensional capabilities of our method.

\begin{figure}[H]
    \centering
    \begin{minipage}{.49\textwidth}
        \centering
\includegraphics[width=0.95\textwidth, trim = {2.5cm 2.5cm 2cm 2cm},clip]{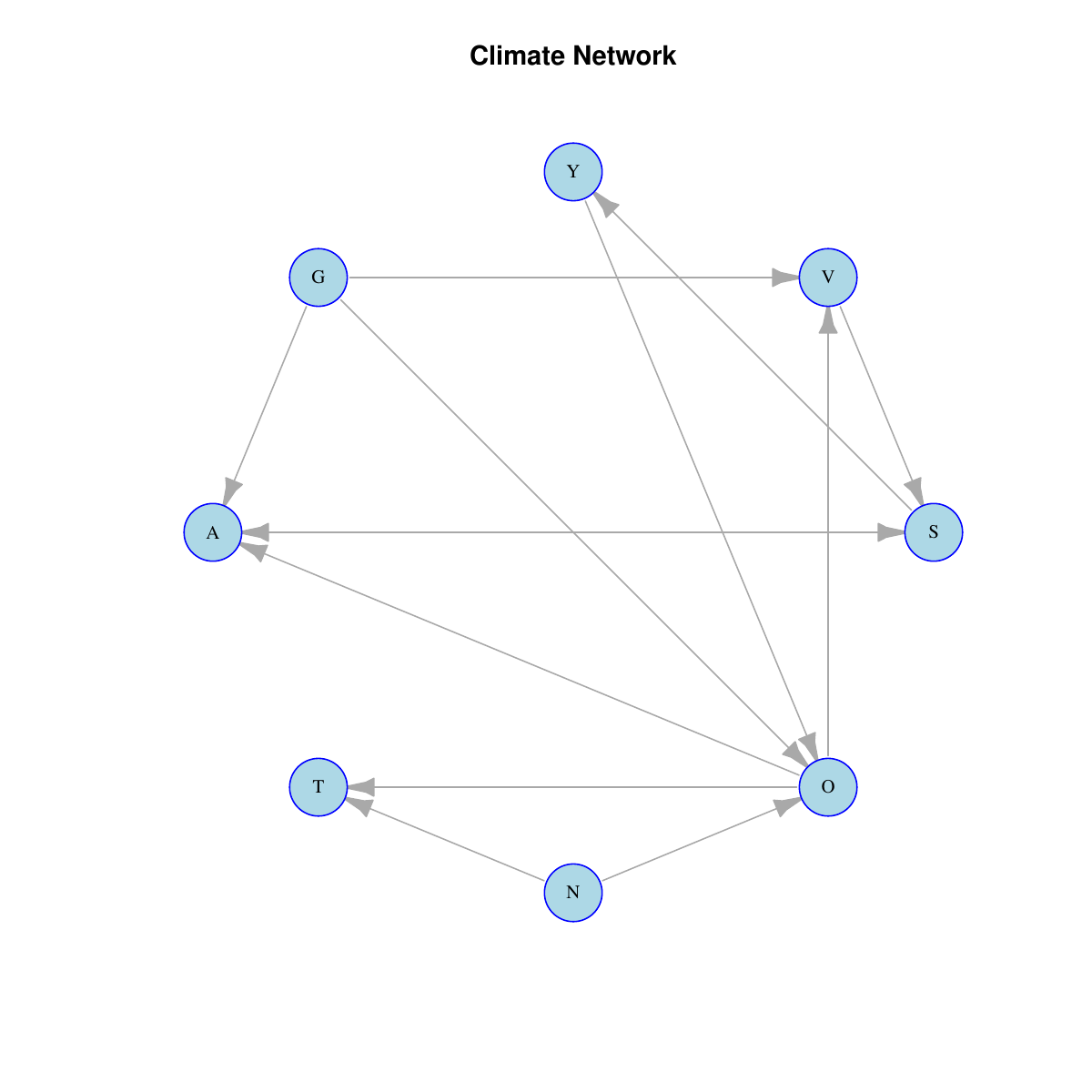}
\caption{Aggregated GHGs, $\alpha=0.1$, $p=10$}
\label{aggresens1}
\end{minipage}%
\begin{minipage}{0.49\textwidth}
\centering
\includegraphics[width=0.95\textwidth, trim = {2.5cm 2.5cm 2cm 2cm},clip]{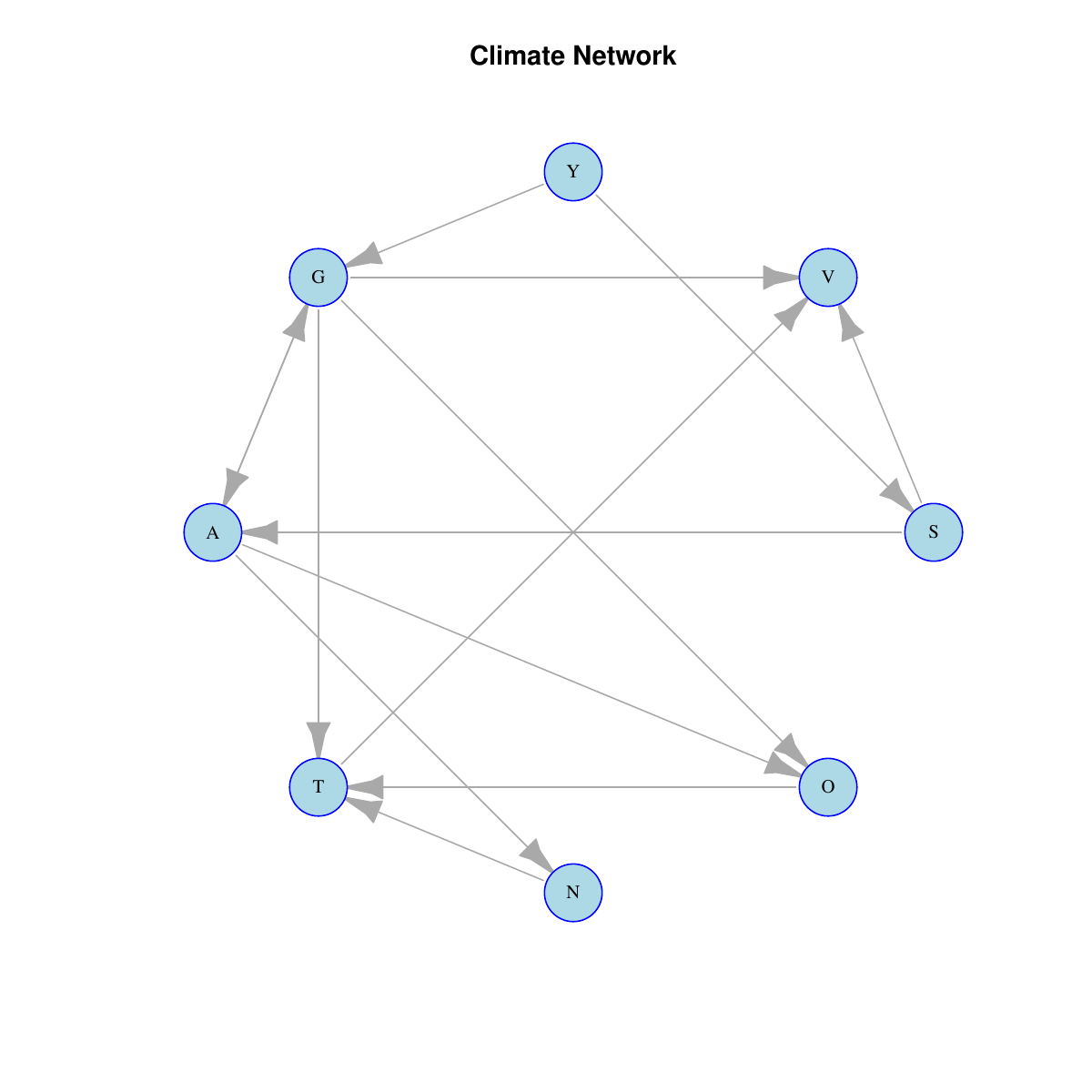}
\caption{Aggregated GHGs,$\alpha=0.1$, $p=30$ }\label{aggresens3}
\end{minipage}
\end{figure}

\begin{figure}[H]
    \centering
    \begin{minipage}{.49\textwidth}
        \centering
\includegraphics[width=0.90\textwidth]{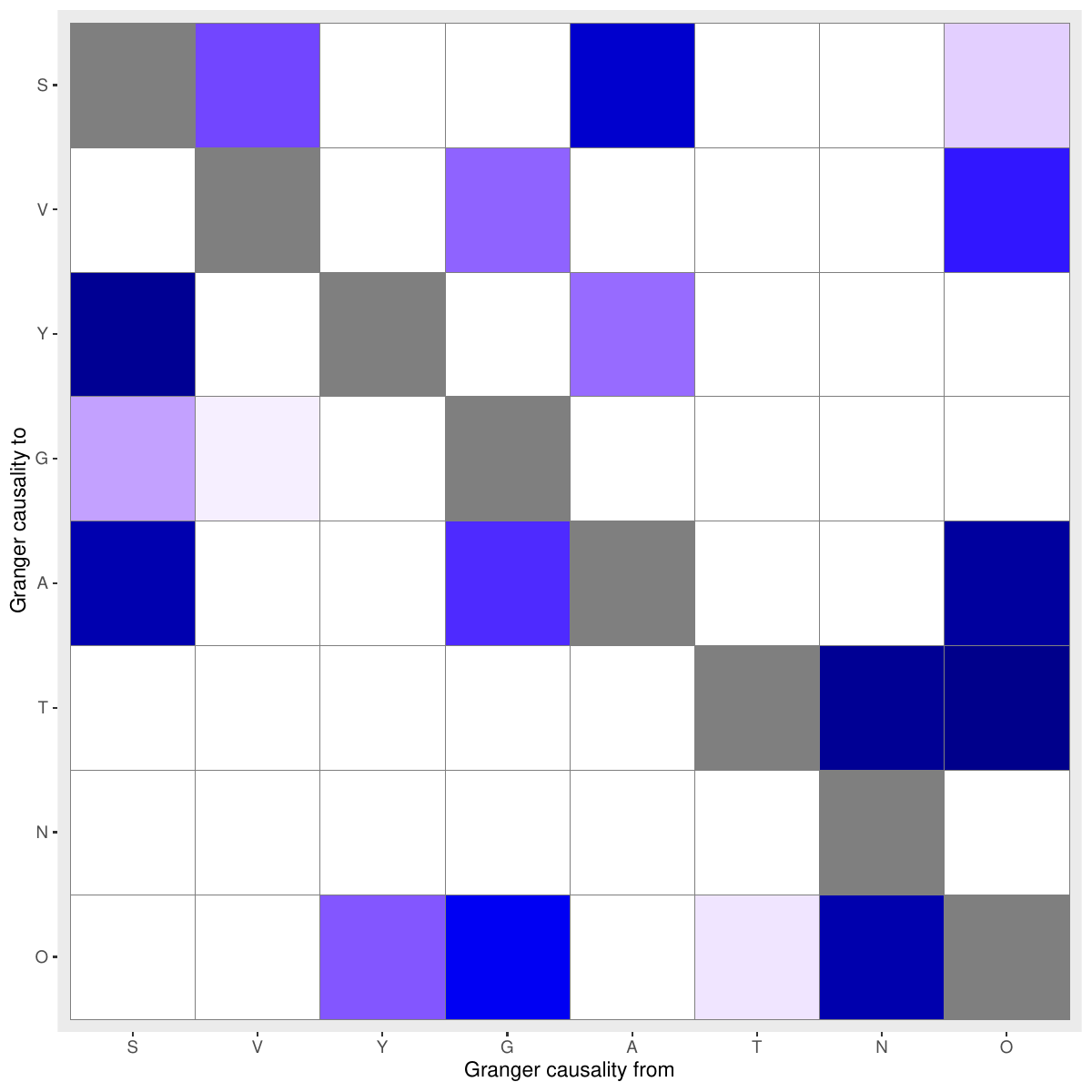}
\caption{\footnotesize{Aggregated GHGs, $p$-values heat-map, $p=10$}}
\label{aggresens4}
\end{minipage}%
\begin{minipage}{0.49\textwidth}
\centering
\includegraphics[width=0.90\textwidth]{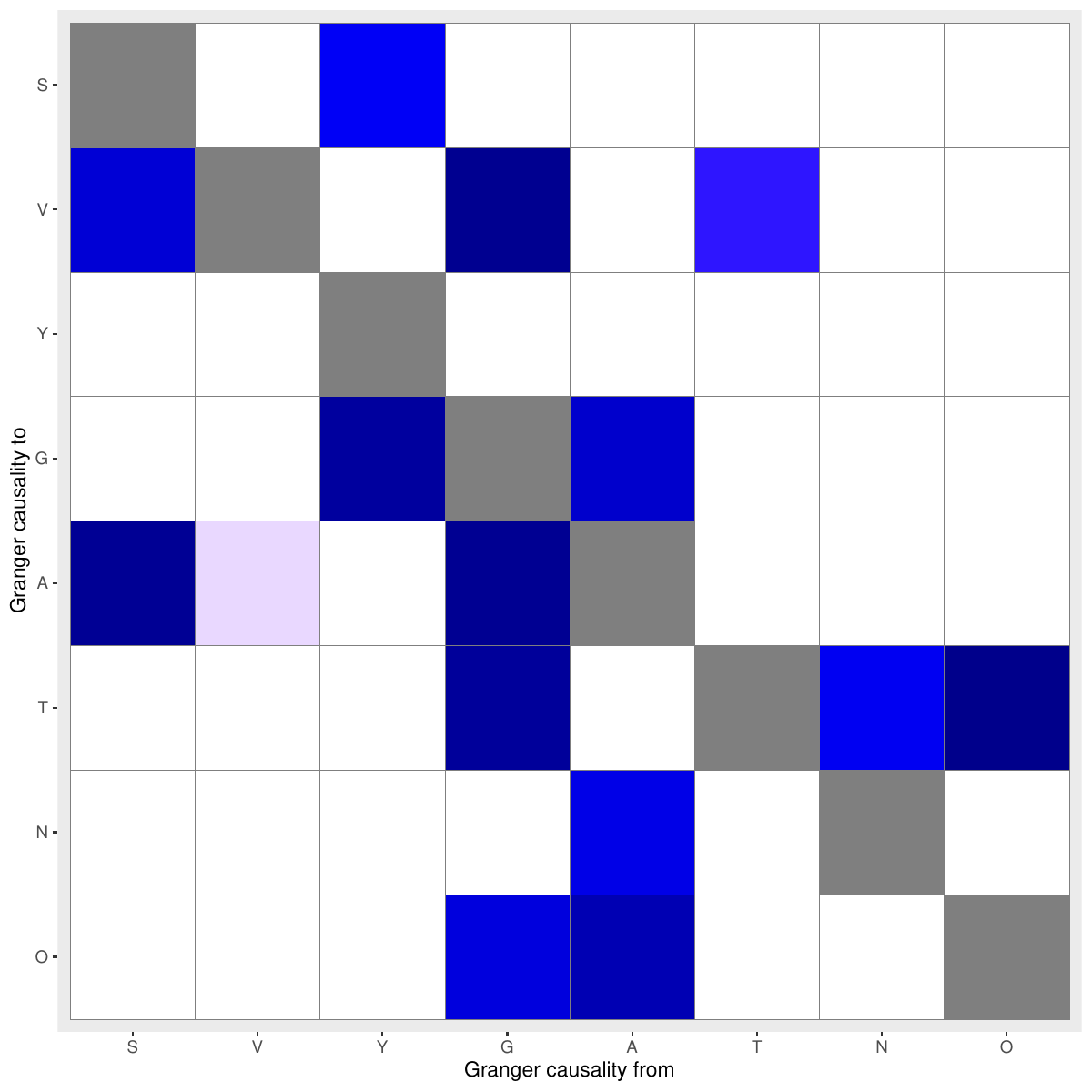}
\caption{\footnotesize{Aggregated GHGs, $p$-values heat-map, $p=30$} }\label{aggresens6}
\end{minipage}
\end{figure}


\begin{figure}[H]
    \centering
    \begin{minipage}{.49\textwidth}
        \centering
\includegraphics[width=0.95\textwidth, trim = {2.5cm 2.5cm 2cm 2cm},clip]{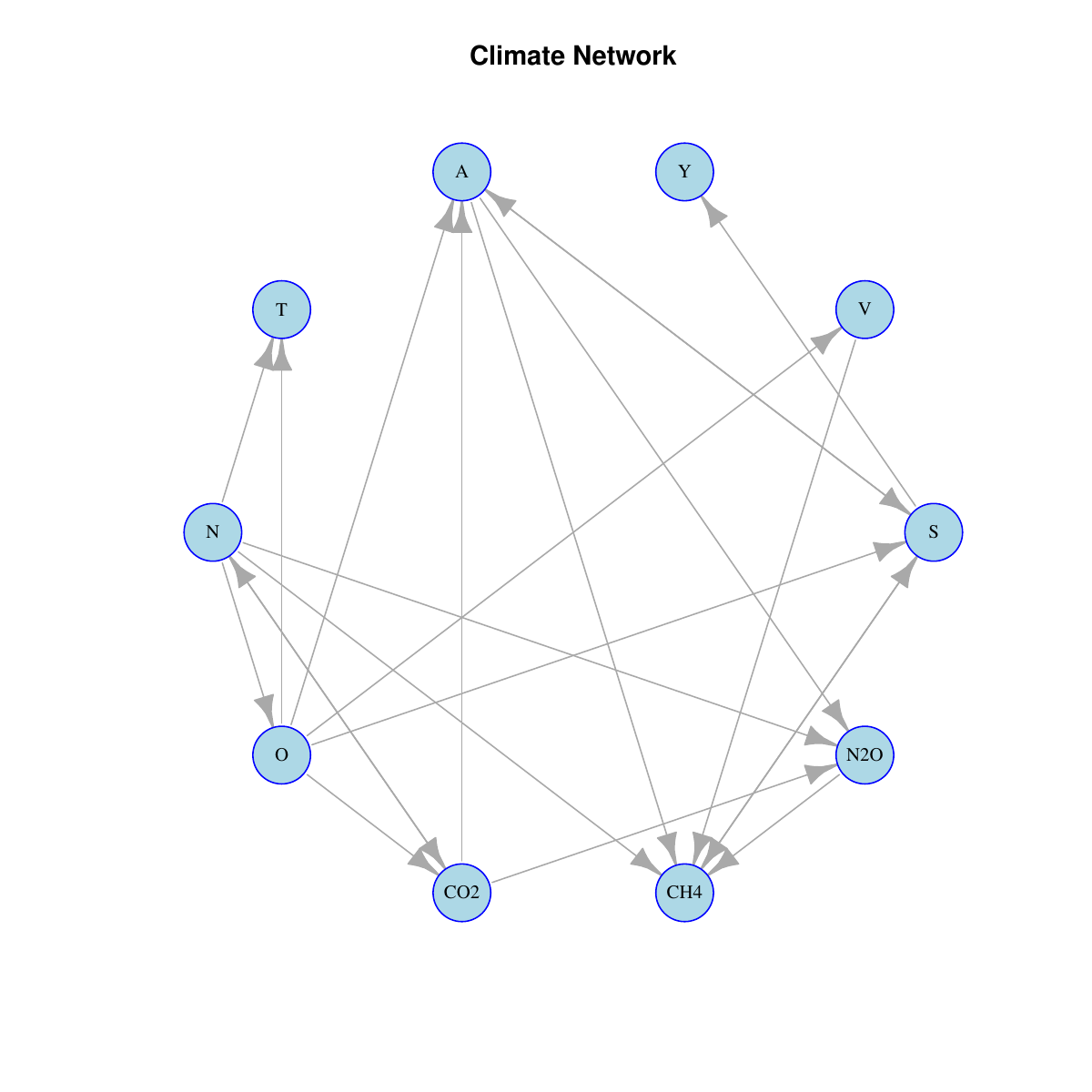}
\caption{Disaggregated GHGs, $\alpha=0.1$, $p=10$}
\label{fig_p10}
\end{minipage}
\begin{minipage}{0.49\textwidth}
\centering
\includegraphics[width=0.95\textwidth, trim = {2.5cm 2.5cm 2cm 2cm},clip]{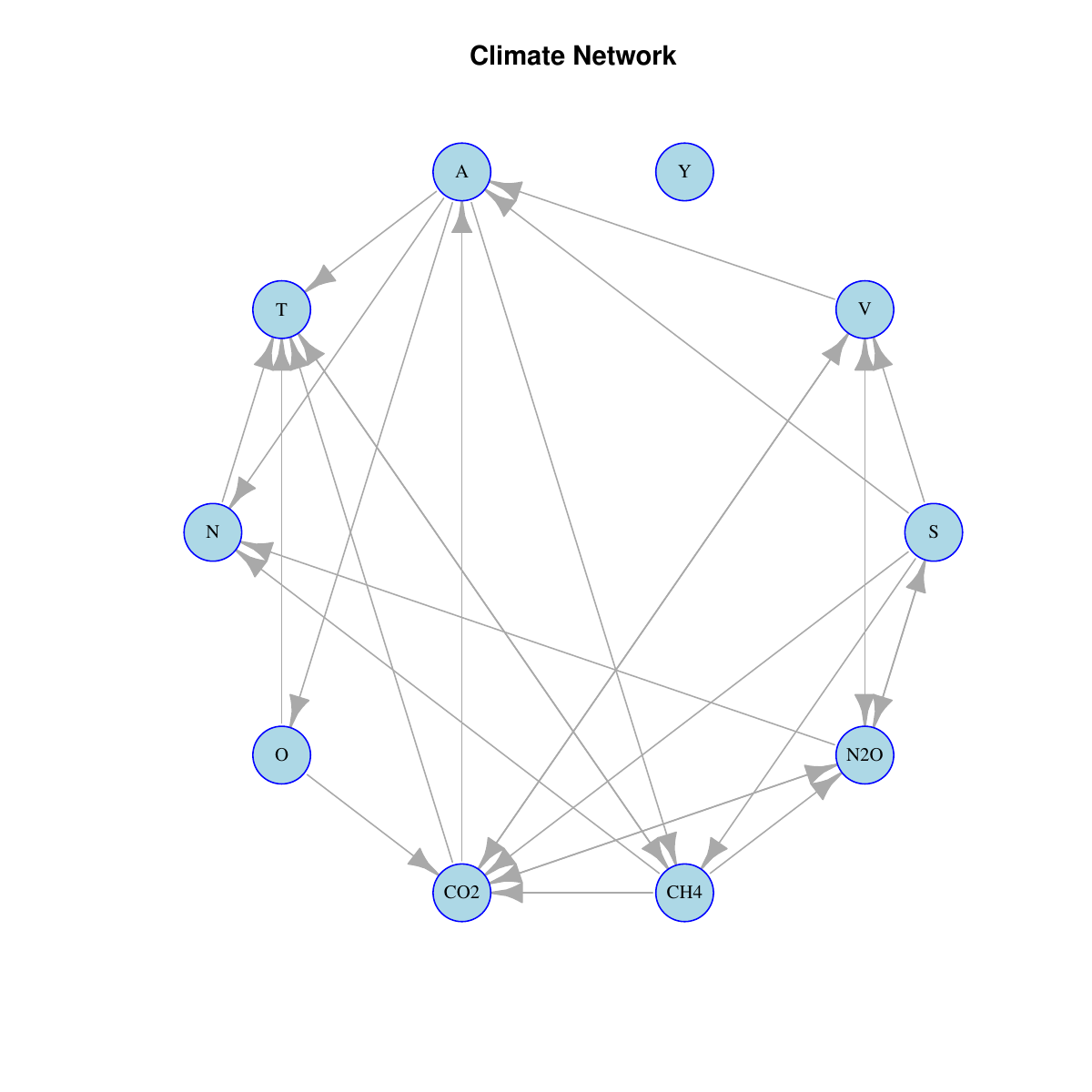}
\caption{Disaggregated GHGs, $\alpha=0.1$, $p=30$ }\label{fig_p30}
\end{minipage}
\end{figure}

\begin{figure}[H]
    \centering
    \begin{minipage}{.49\textwidth}
        \centering
\includegraphics[width=0.95\textwidth]{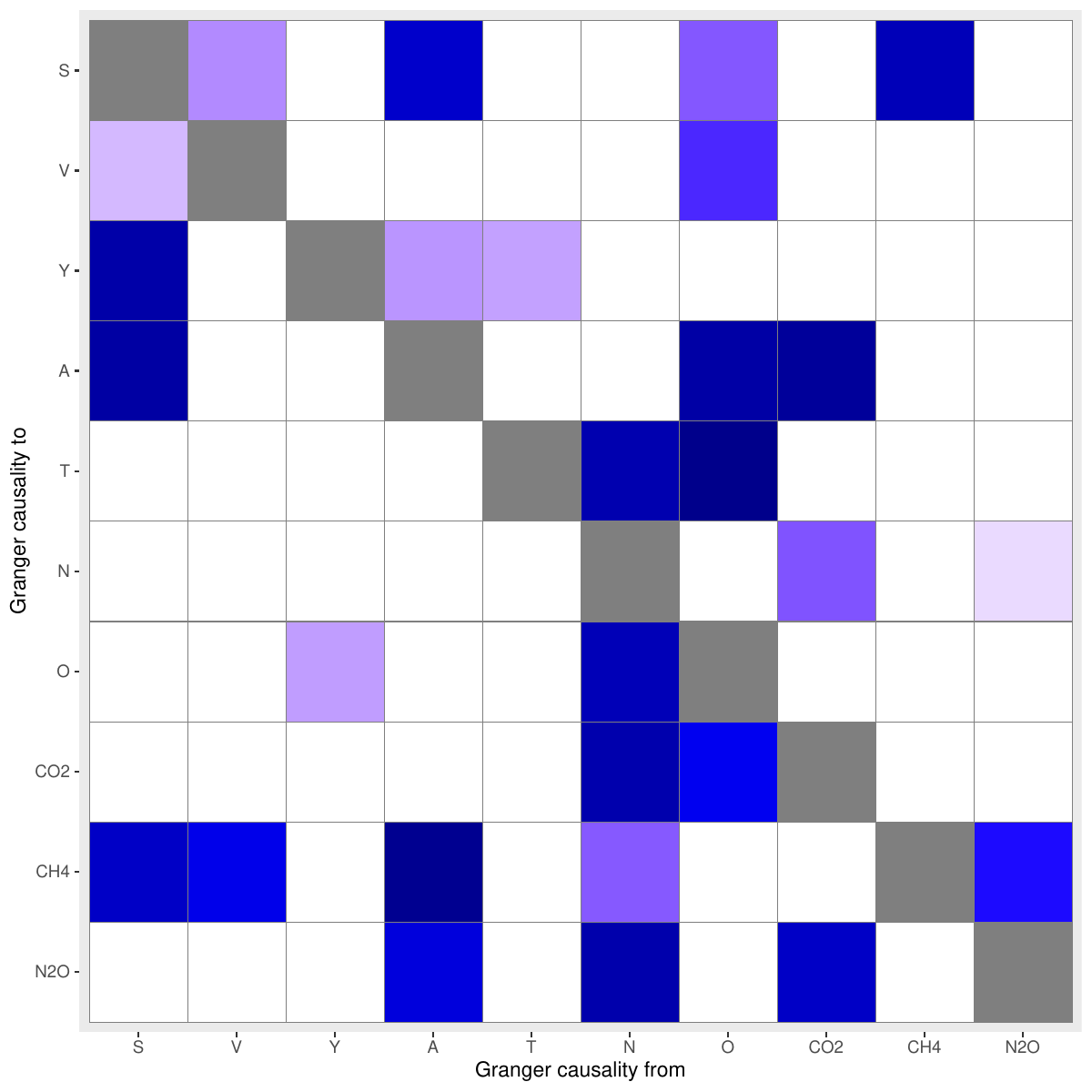}
\caption{\footnotesize{Disaggregated GHGs, $p$-values heat-map, $p=10$}}
\label{fig_p10_1}
\end{minipage}
\begin{minipage}{0.49\textwidth}
\centering
\includegraphics[width=0.95\textwidth]{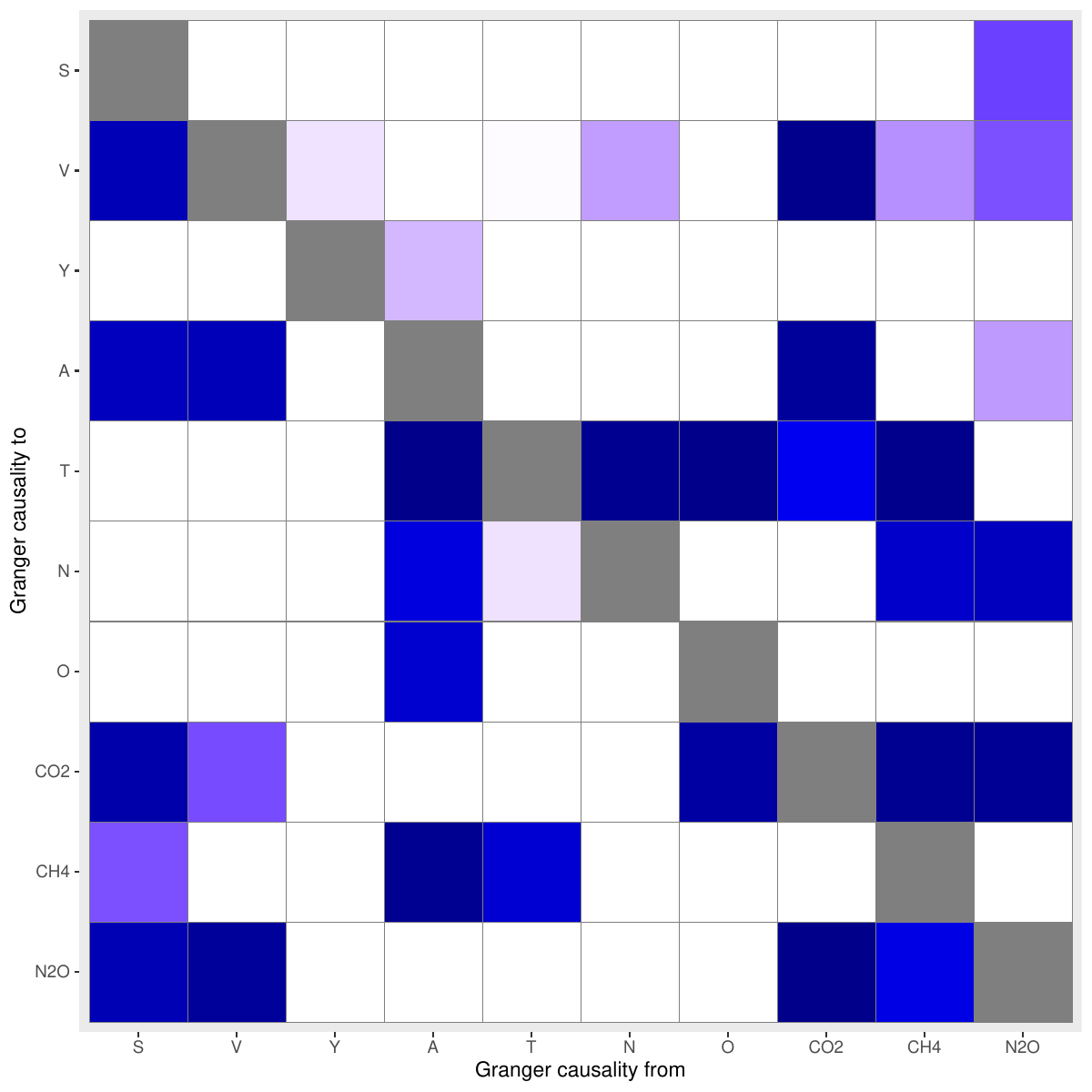}
\caption{\footnotesize{Disaggregated GHGs, $p$-values heat-map, $p=30$}}\label{fig_p30_1}
\end{minipage}
\end{figure}

Figures \ref{aggresens1}, \ref{aggresens3} as well as Figures \ref{fig_p10}, \ref{fig_p30} report the networks estimated respectively with $p=10$ and $p=30$ lags, for both aggregated and disaggregated greenhouse gases. Similarly, Figures \ref{aggresens4}, \ref{aggresens6}, \ref{fig_p10_1}, \ref{fig_p30_1} display the respective $p$-values heat maps. 

From an overall perspective and comparing the results with Figures \ref{fig_miller_4} and \ref{fig_disaggr1} where we had $p=3$, we find a comparable total amount of connections in the systems: in the aggregated case for $p=3$ we found 15, for $p=10$ we found 13, 
and for $p=30$ 14 connections. For the disaggregated case, for $p=3$ we had 25, for $p=10$ now there are 22, 
and for $p=30$ we find 28 connections.
Focusing on those connections involving temperature anomalies ($T$), we find that at all considered lag-lengths ENSO and ocean heat content strongly Granger cause $T$, both at aggregated and disaggregated level.

As expected, accounting for many more lags is particularly beneficial to uncover direct connections between disaggregated greenhouse gases and temperature. We find at 
$p=30$ that both $CO_2$ and $CH_4$ are Granger causal for temperature. We also uncover a direct connection between tropospheric aerosols, surface albedo and temperature at $p=30$. In addition, the dynamics among the greenhouse gases seem to be enhanced by the larger lag-length. At $p=30$, we find that $CO_2$ and $N_2$O have a feedback causal relation and $CH_4$ to be Granger causal both for $CO_2$ and $N_2O$. In particular for $CH_4$, the larger lag-length has uncovered many connections which were previously not visible. 

\section{Discussion and Concluding Remarks}\label{Sec_conclud_rmks} 
We employ high-dimensional Granger causality tests to investigate the connections within climate systems. We particularly focus on the links between radiative forcings and global temperature. Predictive causality in the sense of Granger coupled with a potentially high-dimensional information set are the basis for our causal findings. We employ the Granger causality framework of \citet{hecq2023granger} where honest inference is guaranteed and at the same time no care is needed with respect to the unit-root and cointegration properties of the time series at hand. This is of particular appeal in the case of climate data which are known to contain stochastic trends and exhibit strong persistence.  

We build a dataset containing a set of the most relevant climate time series coupled with GDP, ocean heat content and the El Niño southern oscillation index, for a time frame spanning 1850-2018. We consider two scenarios of increasing dimensionality. We start with a simple system with aggregated greenhouse gases.
We use the lag-augmented post-double-selection procedure described in Section \ref{sec_methodology} on every pair of variables conditional on the remaining ones to obtain a network of Granger causal connections. The system contains GDP in order to account for economic effects on climate variables (and vice-versa). Ocean heat content and the El Niño southern oscillation index are not often included in similar analyses, but recent literature stressed their importance. We indeed find that their inclusion uncovers several important causal connections in the estimated network.

A second, disaggregated setting is also considered, where we decompose the greenhouse gases into their three main components: CO$_2$, CH$_4$ and N$_2$O and we identify indirect causal paths from each of them to temperature, as well as from GDP to temperature. We conclude with a sensitivity analysis in Section \ref{sec_sensitivity}.  There, we manually enlarge the lag-length hoping to uncover connections which might exhibit a slow response to changes in the system. Indeed, we uncover many direct connections between disaggregated greenhouse gases and temperature anomalies, previously masked by the short lag-span considered. 

Building on our findings, there are several directions for future research. First, the recent local projections framework \citep[see e.g.][]{jorda2005estimation, plagborg2019local} could, combined with our approach, provide an impulse response analysis which is common in the VAR literature. This could add interesting findings on the direction and magnitude of responses to shocks in the system. Second, using different data in place of temperature anomalies, such as global mean surface temperature or Earth’s energy imbalance, could further robustify our findings.

\appendix

\section{Block Granger causality}
The outlined methodology in Section \ref{sec_methodology} also allows for testing causality among blocks, rather than just conditional bivariate tests. This implies the possibility of verifying whether a block of variables is Granger causal for another block or alternatively whether a block is Granger causal for a sigle variable. In our context, we could be interested in testing whether the disaggregated greenhouse gases block-Granger causes temperature directly as we observed in Section \ref{subsec_1}. Should be noted that the aggregated series of greenhouse gases in Section \ref{subsec_1} contains also chlorofluorocarbons while in our disaggregated analysis we had to exclude them because of data limitations. 
\begin{figure}[H]
    \centering
\includegraphics[width=0.40\textwidth, trim = {2.5cm 2.5cm 2cm 2cm},clip]{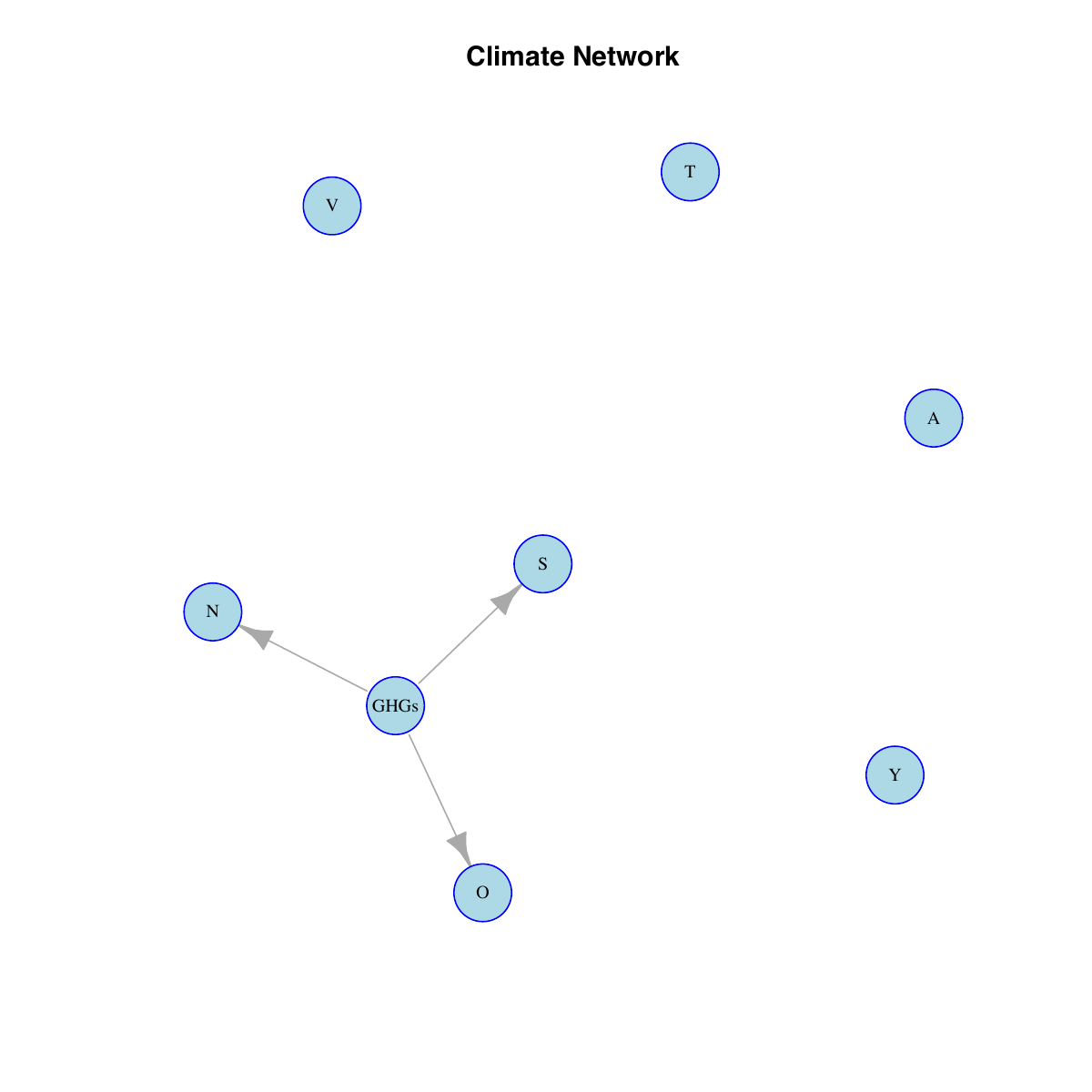}
\caption{Block GHGs connections, $\alpha=0.1$, $p=3$}
\label{fig_Block1}
\end{figure}
In Figure \ref{fig_Block1} we report only the connections between the block of greenhouse gases i.e., $GHGs=(CO_2,N_2O,CH_4)$ and we observe a total of three connections with ocean heat content, ENSO and solar activity, but no direct block-causality with temperature as well as aerosols or GDP. We also observed the direct connection with ocean heat content in Section \ref{subsec_1}, and similarly the connections with solar activity and ENSO, although indirect. 
\begin{figure}
    \centering
\includegraphics[width=0.40\textwidth, trim = {2.5cm 2.5cm 2cm 2cm},clip]{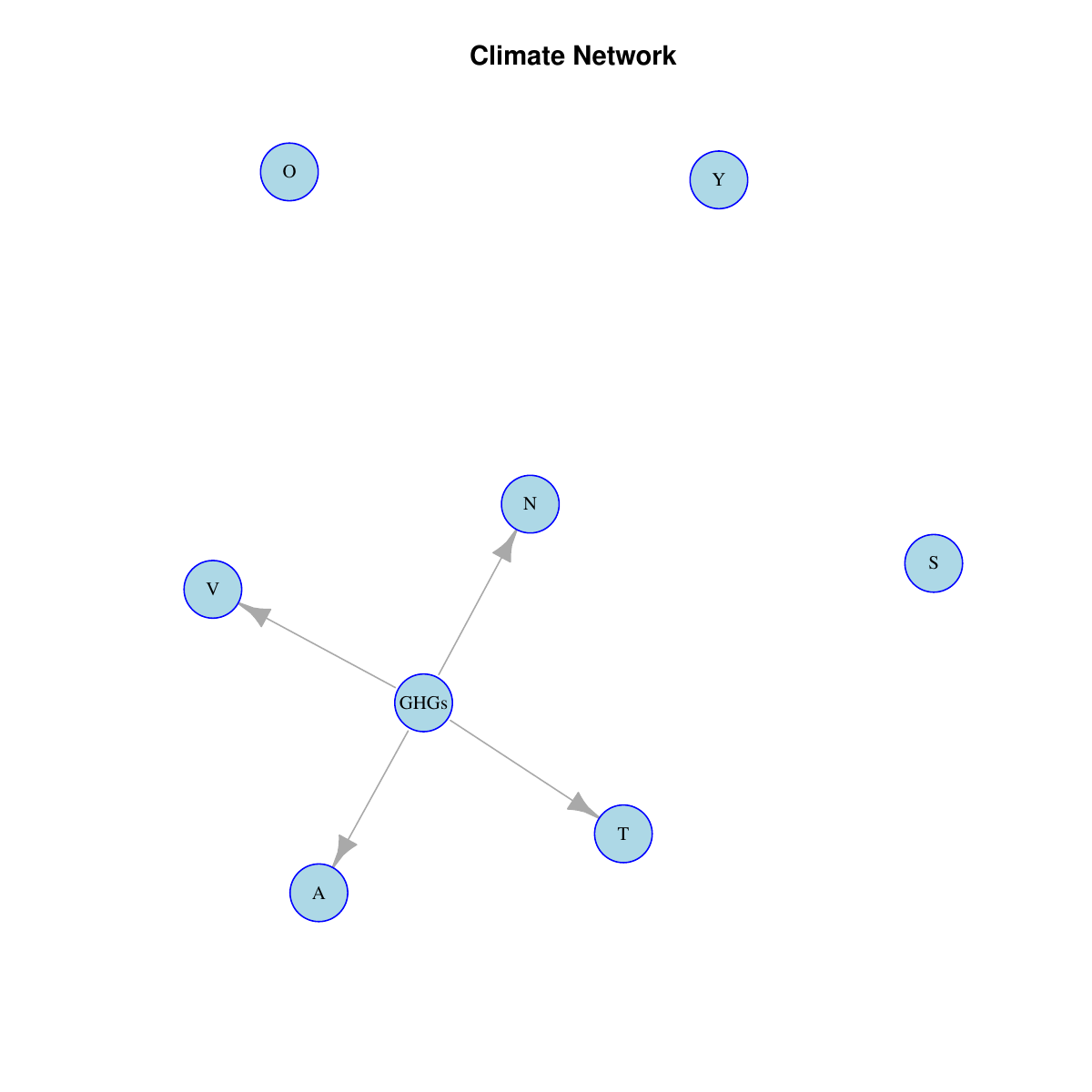}
\caption{Block GHGs connections, $\alpha=0.1$, $p=15$}
\label{fig_Block2}
\end{figure}
In Figure \ref{fig_Block2} we repeat the block-Granger causality analysis as at the end of Section \ref{subsec_3} using $p=15$ lags. We now find the greenhouse gases considered i.e., CO$_2$, N$_2$O and CH$_4$, when considered as a block, they do Granger cause temperature as expected, and as already observed in Section \ref{subsec_3}.

\bibliography{literature}
\end{document}